\documentclass[twocolumn,showpacs,preprintnumbers,amsmath,amssymb,prb]{revtex4-1}

\usepackage{amsmath,amssymb}
\usepackage{epsfig}
\usepackage{graphicx}
\usepackage{bm}

\renewcommand{\v}{{\bf v}}
\renewcommand{\j}{{\bf j}}
\renewcommand{\k}{{\bm{k}}}

\newcommand{\re}{{\rm{Re}}}
\newcommand{\im}{{\rm{Im}}}

\def\gsim{\lower.35em\hbox{$\stackrel{\textstyle>}{\textstyle\sim}$}}

\begin{document}
\title{Collective magnetic excitations in AA- and AB-stacked graphene bilayers}

\author{M. S\'anchez S\'anchez$^{1,2}$, G. G\'omez-Santos$^2$, and T. Stauber$^{1,3}$}
\affiliation{
$^1$ Materials Science Factory, Instituto de Ciencia de Materiales de Madrid, CSIC, E-28049 Madrid, Spain\\
$^2$Departamento de F\'{\i}sica de la Materia Condensada, Instituto Nicol\'as Cabrera and Condensed
Matter Physics Center (IFIMAC), Universidad Aut\'onoma de Madrid, E-28049 Madrid, Spain\\
$^{3}$ Theoretical Physics III, Center for Electronic Correlations and
Magnetism, Institute of Physics, University of Augsburg, D-86135
Augsburg, Germany}
\date{\today}

\begin{abstract}
We discuss novel transverse plasmon-polaritons that are hosted by AA- and AB-stacked bilayer graphene due to perfect nesting. They are composed of oscillating counterflow currents in between the layers, giving a clear interpretation for these collective modes as magnetic excitations carrying magnetic moment parallel to the planes. For AA-stacked bilayer graphene, these modes can reach zero frequency at the neutrality point and we thus predict a symmetry broken ground-state leading to in-plane orbital ferromagnetism. Even though it could be hard to detect them in real solid-state devices, these novel magnetic plasmons should be observable in artificial set-ups such as optical lattices. Also, our results might be relevant for magic angle twisted bilayer graphene samples as their electronic properties are mostly determined by confined AA-stacked regions.
\end{abstract}

\maketitle

\section{Introduction}
Plasmonics in graphene-based materials has attracted much attention since the wavelength of light can be reduced by several orders of magnitude governed by the inverse fine-structure constant $\alpha^{-1} \sim137$.\cite{Chen12,Fei12,Grigorenko12,Bludov13,Stauber14,Peres16,Low17} This strong confinement further gives rise to various applications such as  perfect absorption,\cite{Koppens11} the sensoring of biochemical molecules, \cite{Rodrigo15} plasmon-induced transparency\cite{Baqir19} or plasmon-enhanced chirality.\cite{Stauber20} Also other two-dimensional systems have shown interesting plasmonic properties such as molybdenum disulfide,\cite{Scholz13} black phosphorous,\cite{Low14} or general van der Waals materials.\cite{Basov16} 

Plasmonics is usually based on longitudinal or transverse-magnetic (TM) plasmons which consist of collective density oscillations collinear to the propagation direction. These excitations are strongly confined due to the enhanced Coulomb coupling between the electromagnetic field and the charge carrier density. On the other hand, transverse or transverse-electric (TE) plasmons can also exist in graphene which consist of collective current oscillations that are perpendicular to the propagation direction.\cite{Mikhailov07} These excitations are closely pinned to the light-cone and thus weakly confined, which allows for their detection in Otto configuration.\cite{Ramos-Mendieta14,Menabde16} Negative refractive index environments can help to enhance their confinement, \cite{Zhang20,Reserbat-Plantey21} and magnetically biased graphene-ferrite structures give rise to non-reciprocal plasmons.\cite{Chamanara16} Also, in AB-stacked bilayer graphene the transverse modes are considerably more confined than in its monolayer counterpart.\cite{Jablan11} The non-linear response of transverse plasmons was analyzed in Ref. \onlinecite{Andreeva18}. 

Transverse plasmons in single layer graphene arise from interband transitions close to the absorption threshold and it is difficult to visualize their magnetic character.\cite{Principi09,Stauber10,Angel13} This might further be a reason for their fragile nature, i.e., their ultrahigh refractive index sensitivity.
\begin{figure}[h]
    \centering
    \includegraphics[width=.9\linewidth]{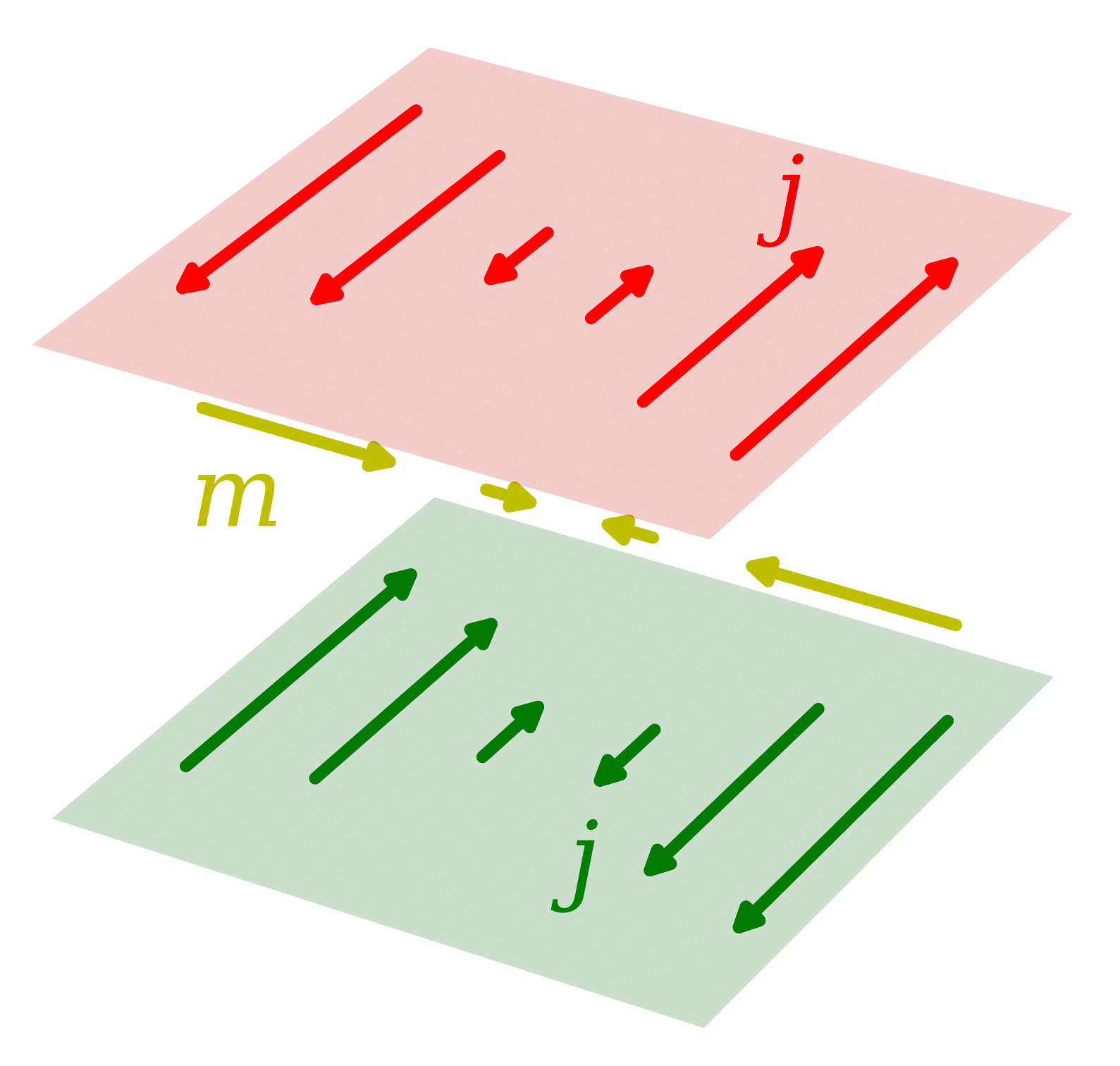}
    \caption{Sketch of the transverse antisymmetric plasmons in bilayer graphene. The loop currents $j$ create an in-plane magnetic moment, $m$, between the layers. In particular, we are interested in the long-wavelength limit $q\to0$ where a uniform, in-plane,  oscillating magnetic  moment  emerges  out  of  transverse  (chargeless)  currents which are opposite in each layer.}
    \label{fig:my_label}
\end{figure}

Here, we will study the antisymmetric transverse mode of AA or AB-stacked graphene bilayers that has not been mentioned in the literature, so far.\cite{Basov21} They are considerably more stable as they arise from a resonant "magnetic" absorption, absent in uncoupled bilayers which were discussed in Ref. \onlinecite{StauberPRB12}. Further, the antisymmetric counterflow leads to perfect screening outside the sample for $qa\to0$ such that the plasmon dispersion does not depend on the two external, possibly different dielectric environments, contributing to additional stability. Finally, these antisymmetric modes have a clear physical interpretation and correspond to loop currents around the layers resulting in an in-plane magnetic moment density, see Fig. \ref{fig:my_label}. Remarkably, for AA-stacked graphene these modes can also exist at zero frequency and we thus predict a magnetic instability and spontaneous symmetry breaking at zero temperature, giving rise to orbital in-plane ferromagnetism.

The paper is organized as follows: in Sec. \ref{ElectromagneticResponse}, we discuss the electromagnetic response in a bilayer and define the plasmonic excitations within the random-phase approximation (RPA). We also comment on the effect of an inhomogeneous dielectric environment for the transverse (magnetic) modes. In Sec. \ref{CurrentResponse}, we discuss the magnetic response of two typical graphene bilayers that are AA and AB-stacked and give numerical estimates for the resonant frequency. We finally discuss the stability in AA-stacked bilayer graphene in Sec. \ref{MagneticInstability} and close with conclusions.

\section{Electromagnetic Response in Bilayers}
\label{ElectromagneticResponse}
Following Ref. \onlinecite{Stauber18}, the most general (bare) in-plane current response in isotropic (and non-chiral) bilayer systems can be written as
\begin{align}
\left(
\begin{array}{c}
j_\nu^1\\
j_\nu^2
\end{array}\right)=-\left(
\begin{array}{cc}
\chi_{11}&\chi_{12}\\
\chi_{21}&\chi_{22}
\end{array}\right)\left(
\begin{array}{c}
A_\nu^1\\
A_\nu^2
\end{array}\right)\;,
\end{align}
where $A_\nu^\ell$ denotes the gauge field acting on layer $\ell=1,2$ and $\nu=l,t$ defines the longitudinal and transverse channels, which are decoupled. Further, $\chi_{\ell,\ell'}=\ll j_\nu^\ell,j_\nu^{\ell'}\gg$ is the current-current response with respect to layer $\ell$ and $\ell'$. Notice that the polarization subscript $\nu$ is suppressed in $\chi_{n,m}$ since both polarizations coincide in the $q\to0$-limit.

In a homogeneous dielectric environment, the propagators of the gauge fields with in-plane momentum $q$ and frequency $\omega$ are given by
\begin{align}
    {\bf \mathcal{D}_\nu} = d_\nu \begin{pmatrix}
    1 && e^{-q'a} \\ e^{-q'a} && 1 \\
    \end{pmatrix},
\end{align}
with $a$ the distance between layers, $q'=\sqrt{q^2-\epsilon\mu\omega^2/c^2}$, $d_l=\frac{q'}{2\epsilon_0\omega^2}$ and $d_t=-\frac{1}{2\epsilon_0c^2q'}$.\cite{Stauber12} Time-reversal symmetry sets $\chi_{12} = \chi_{21}$, and in the following, we further assume that the exchange of layers is also a symmetry leading to $\chi_{11} = \chi_{22}$. The collective excitations (plasmons) are then defined in the random-phase approximation by\cite{giuliani05}
\begin{align}
\textrm{det}({\bf 1 - \chi \mathcal{D}_\nu}) = 0\;,
\end{align}
where $\chi=\chi_{\ell,\ell'}$ denotes the $2\times2$-response matrix.

In terms of bonding and anti-bonding modes, we have $\chi_{\pm} = \chi_{11} \pm \chi_{12}$, $d_{\nu \pm} = d_{\nu}(1 \pm \exp(-q'a))$, and plasmons are the zeroes of the effective dielectric functions:
\begin{align}
\label{PlasmonCondition}
1-d_{\nu\pm}\chi_\pm=0.
\end{align}
The transverse field propagator is purely imaginary in the $qa\to0$ limit,  making any current excitation to leak out to the propagating electromagnetic (EM) field for a single layer. This is avoided in the double layer by the combination $1-\exp(-2q'a)$ in $d_{t-}$, making the effective propagator for the antisymmetric mode real to lowest order in $q$- Leaking to the EM field thus only occurs in higher orders.  Of course,  this effective real magnetic coupling is still very small, but the condition for the self-sustained collective modes is guaranteed by the divergent nature of the current response $\chi_-$ near sharp features of the spectrum as we will show below.

The response function $\chi_-$ refers to the counterflow that leads to an in-plane magnetic moment. The intrinsic excitations are thus collective magnetic dipole oscillations or magnetic plasmons - contrary to the conventional plasmons which  are collective electric dipole oscillations. Nevertheless, the resonance condition (\ref{PlasmonCondition}) is usually not fulfilled due to the weak magnetic coupling and a true magnetic plasmon cannot be formed. 

In the next Section we will show that in typical graphene bilayers with AA and AB-stacking, the resonance condition can always be reached due to "perfect nesting" close to a frequency $\omega_0$, where the real part of $\chi_-$ diverges as $(\omega-\omega_0)^{-1}$. For extremely clean samples, collective modes are thus guaranteed by the divergent nature of the current response $\chi_-$ near sharp features of the spectrum and we predict genuine magnetic excitations close to the resonance condition. 

Interestingly, in an inhomogeneous environment, i.e., if the bilayer is surrounded by two dielectrics with permittivities $\epsilon_1$ and $\epsilon_2$, and permeabilities $\mu_1$ and $\mu_2$, the final result for the antisymmetric modes does not change in the long-wavelength limit due to perfect screening. The antisymmetric eigenvalue of $\bf{1}-\bf{\chi \mathcal{D}_t}$ then reads (see Appendix B)
\begin{align}
    \lambda^- &= 1 + \frac{\mu_0 a}{2}(\chi_{11} - \chi_{12})\left( 1 + \mathcal{O}(q'a)\right)\;.
\end{align}
The magnetic mode, alternatively defined by $\re\lambda^-=0$, is therefore essentially independent of the surrounding dielectric media. On the other hand, the symmetric plasmonic mode does depend on the environment, and the existence of excitations closely pinned to the light cone will strongly depend on the refractive index difference, $n_1 - n_2$.\cite{Kotov13}

\section{Graphene bilayers and current response}
\label{CurrentResponse}
The calculation of $\chi_{11}$ and $\chi_{12}$ only requires the knowledge of the underlying Hamiltonian. We consider the general low-energy Hamiltonian of graphene bilayers\cite{Castro09,McCann13,Rozhkov16}
\begin{align}
H=\sum_\k H_\k\;,\;\text{with}\;
H_\k=\left(
\begin{array}{cc}
H_\k^0&U^\dagger\\
U&H_\k^0
\end{array}\right)\;,
\end{align}
where $H_\k^0=\hbar v_F{\bm\sigma}\cdot\k$ is the single layer graphene Dirac Hamiltonian acting on states of Bloch momentum $\k$. The $2\times2$-matrix $U$ denotes the interlayer coupling, characterized by the interlayer hopping matrix element $t\sim0.33$eV.

We are interested in the generalized current-current response given by the Kubo formula\cite{giuliani05}
\begin{align}
\chi^{i,i'}_{\ell,\ell'}=\frac{e^2g_sg_v}{A}\sum_{n,m}\frac{f_{n}-f_{m}}{\hbar\omega-(\epsilon_{m}-\epsilon_{n})+i\eta}v_{n;m}^{i,\ell}v_{m;n}^{i',\ell'}\;,
\label{kubo}
\end{align}
where $i,i'=x,y$ denote the spatial direction, $\ell,\ell'=1,2$ the different layers and $n,m$ the set of quantum numbers; and $g_s = 2, g_v=2$ are the spin and valley degeneracy respectively. The charge of the electron is $e$, $f_n$ is the Fermi-Dirac distribution, and $v_{n;m}=\langle n|{\hat v}|m\rangle$ are the matrix elements of the general velocity operator ${\hat v}$, which in the local approximation ($q=0$) couples states with the same momentum, $\k$. 

The "magnetic" velocity can be defined by the counterflow between the two layers and the magnetic moment is thus proportional to this velocity and the area (distance) between the two layers. For the external gauge field in $i$-direction, the magnetic response is thus defined by $\chi_-=\chi_{11}^{ii}-\chi_{12}^{ii}$, but we will suppress the super-index in the following.
\begin{figure}[h]
    \centering
    \includegraphics[width=.96\linewidth]{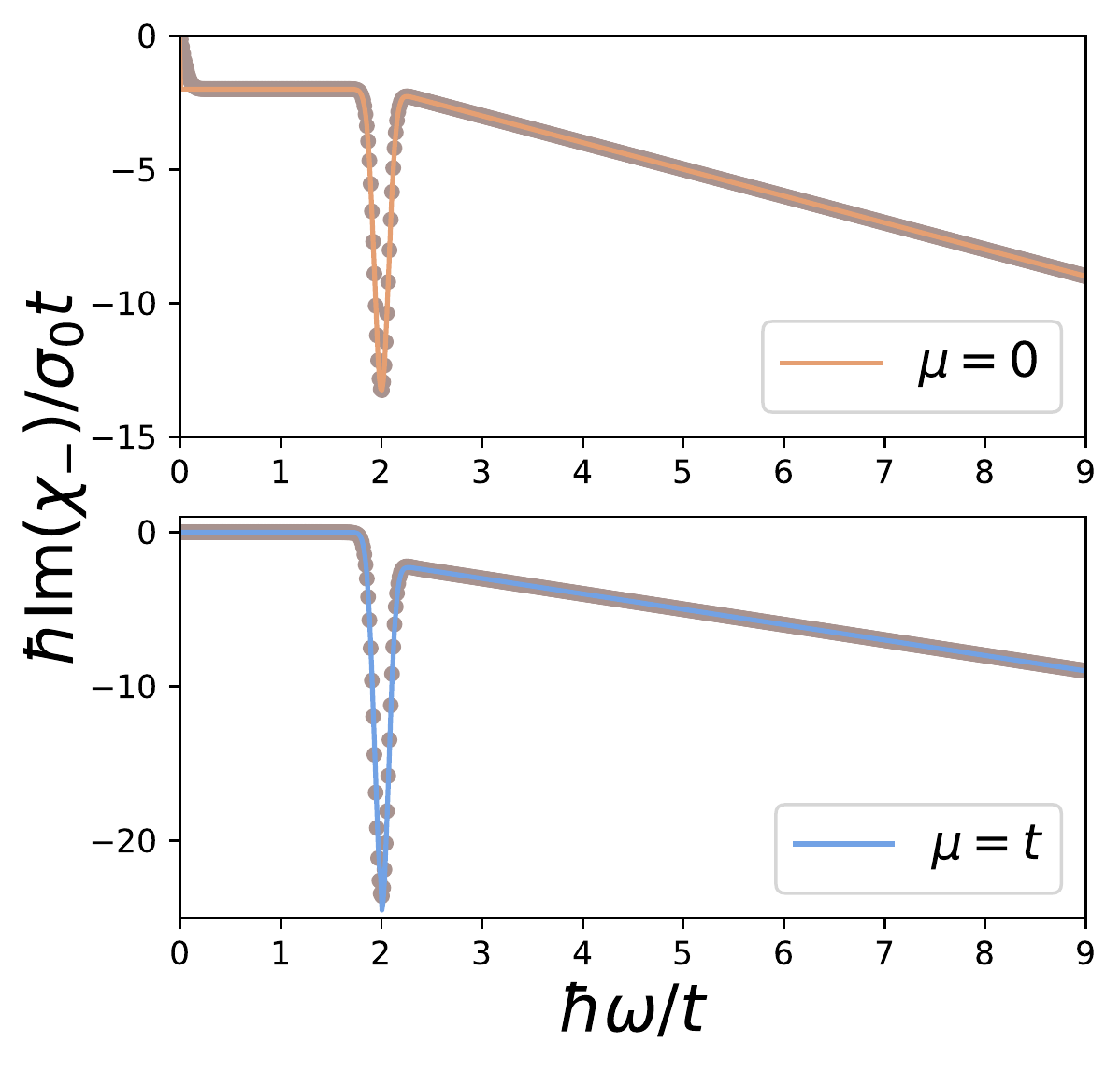}
    \includegraphics[width=.96\linewidth]{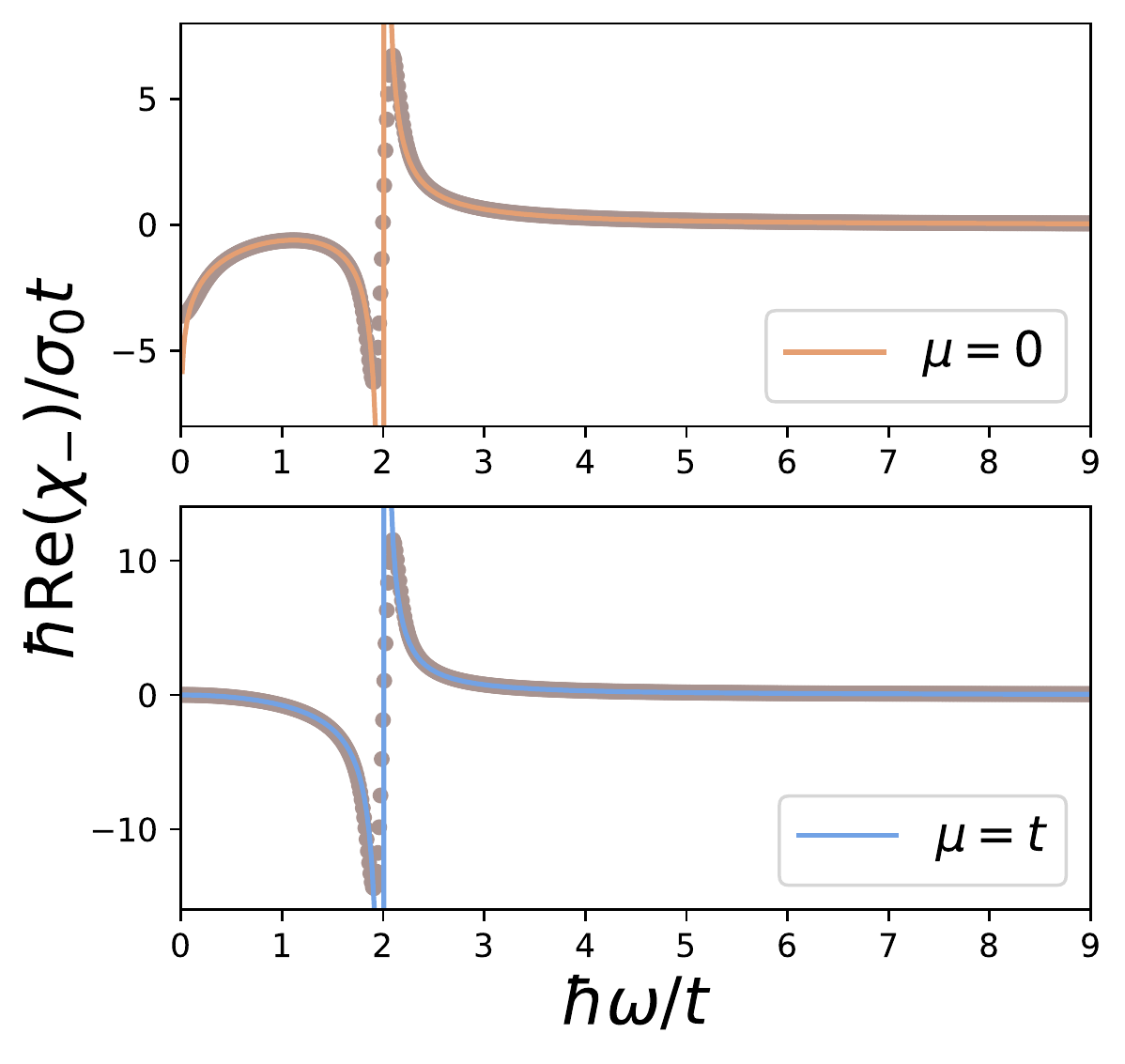}
    \caption{Magnetic response function for $AA-$stacked bilayer graphene as function of the frequency, for $\mu=0$ and $\mu = t$. Top: imaginary part of the magnetic response, $\im \chi_{-}$, in units of $\sigma_0 t /\hbar$, where $\sigma_0 = e^2/4\hbar$ is the universal conductivity of graphene. 
    Bottom: real part of the magnetic response function, $\re \chi_{-}$, in units of $\sigma_0 t /\hbar$. The dots are the results of the numerical computations, and the solid lines are the formulas derived in Appendix A. The width of the delta function has been tuned to match the numerical results for which $\eta=0.1t$ was used.}
    \label{chi}
\end{figure}

The magnetic susceptibility to an in-plane magnetic field can be obtained from the magnetic response. If $m_\parallel$ and $B_\parallel$ are moment and field, then writing $m_\parallel=\chi_\parallel B_\parallel$, one has $2\chi_\parallel=-a^2\chi_-$.  Notice that the emergence of a magnetic mode requires paramagnetism for which $\chi_\parallel$ becomes positive. 

\subsection{AA-stacked bilayer graphene}
With $U=t\left(
\begin{array}{cc}
1&0\\
0&1
\end{array}\right)$, we describe AA-stacked graphene whose optical conductivity has been discussed in Ref. \onlinecite{Tabert12} and the plasmonic properties in Ref. \onlinecite{Roldan13}.

The eigenstates are the bonding and anti-bonding states of single-layer graphene. The eigenenergies are given by $E_{k,n_1,n_2}=n_1t+n_2\hbar v_Fk$ and the eigenvectors by $\psi_\k^{n_1,n_2}=\frac{1}{2}(n_1,n_1n_2e^{i\varphi_\k},1,n_2e^{i\varphi_\k})^T$ where $n_1,n_2=\pm$ and $\varphi_\k$ denotes the angle that $\k$ forms with the $x$-axis. The velocity operator of layer $\ell$ is given by $\hat\v_\ell={\bm\sigma}\otimes{\bm1}_\ell$ where ${\bm1}_\ell$ performs a projection onto states of sublayer $\ell$ and ${\bm\sigma}$ are the Pauli matrices of the pseudospin variables. 

The "electric"/"magnetic" excitations couple to  the  total/counterflow current $\hat\j_{\pm}=-e\hat\v_{\pm}=-e(\hat\v_1\pm\hat\v_2)$ and the relevant matrix element is
\begin{align} 
v^{\pm}_{\k,n_1,n_2;\k,m_1,m_2}=(n_1m_1\pm 1)\frac{v_F}{4}\left(m_2e^{i\varphi_{\k}}+n_2e^{-i\varphi_\k}\right)\;.
\end{align}
This means that only transitions with $n_1m_1=-1$ yield a finite contribution to the imaginary part of the "magnetic" response.
Interestingly, for $\hbar\omega=2t$ there is a perfect matching that gives rise to a delta-function and which is only present in the "magnetic absorption" (see Fig. \ref{bands}). The resonant contribution of the magnetic response $2\chi_-=\ll j_{-}^xj_{-}^x\gg$ 
at $T=0$ reads 
\begin{align}
\im\chi_{-}^{nested}(\omega)=&-\frac{e^2g_sg_v}{16\hbar^2}\delta(\hbar \omega - 2t) \\ \nonumber
\times &\Big[ 2(\mu^2 + t^2) \theta(t-|\mu|) + 4t|\mu|\theta(|\mu| -t) \Big] \;.
\end{align}

The real part is given by the Kramers-Kronig relation
\begin{align}
\re\chi_{-}(\omega)=\frac{1}{\pi}\int_0^\infty d\omega'\im\chi_{-}(\omega')\frac{2\omega'}{(\omega')^2-\omega^2}\;,
\end{align}
and the main contribution from the perfect nesting condition is thus given by
\begin{align}
\re\chi_{-}^{nested}&=-\frac{1}{\pi}\frac{e^2g_sg_v}{8\hbar^2}\frac{2t}{(2t)^2-(\hbar\omega)^2} \\\nonumber&\times\big[2(\mu^2 + t^2) \theta(t-|\mu|) + 4t|\mu|\theta(|\mu| -t) \big]\;.
\end{align}

In Fig. \ref{chi}, we show the real and imaginary part of the full magnetic response as discussed in Appendix A for $\mu=0$ and $\mu=t$. The delta-function of the imaginary part of the analytical solution (full line) is broadened by the same value ($\eta = 0.1t$) as used in the numerical solution (dots). The singularity in the real part of the analytical solution, however, is plotted for the clean system which is algebraic for both chemical potentials at $\hbar\omega=2t$. For $\mu=0$, there is a logarithmic singularity at $\omega=0$ which is responsible for the symmetry-broken ground-state.

\subsection{AB-stacked bilayer graphene}

\begin{figure}[b]
    \centering
    \includegraphics[width=.95\linewidth]{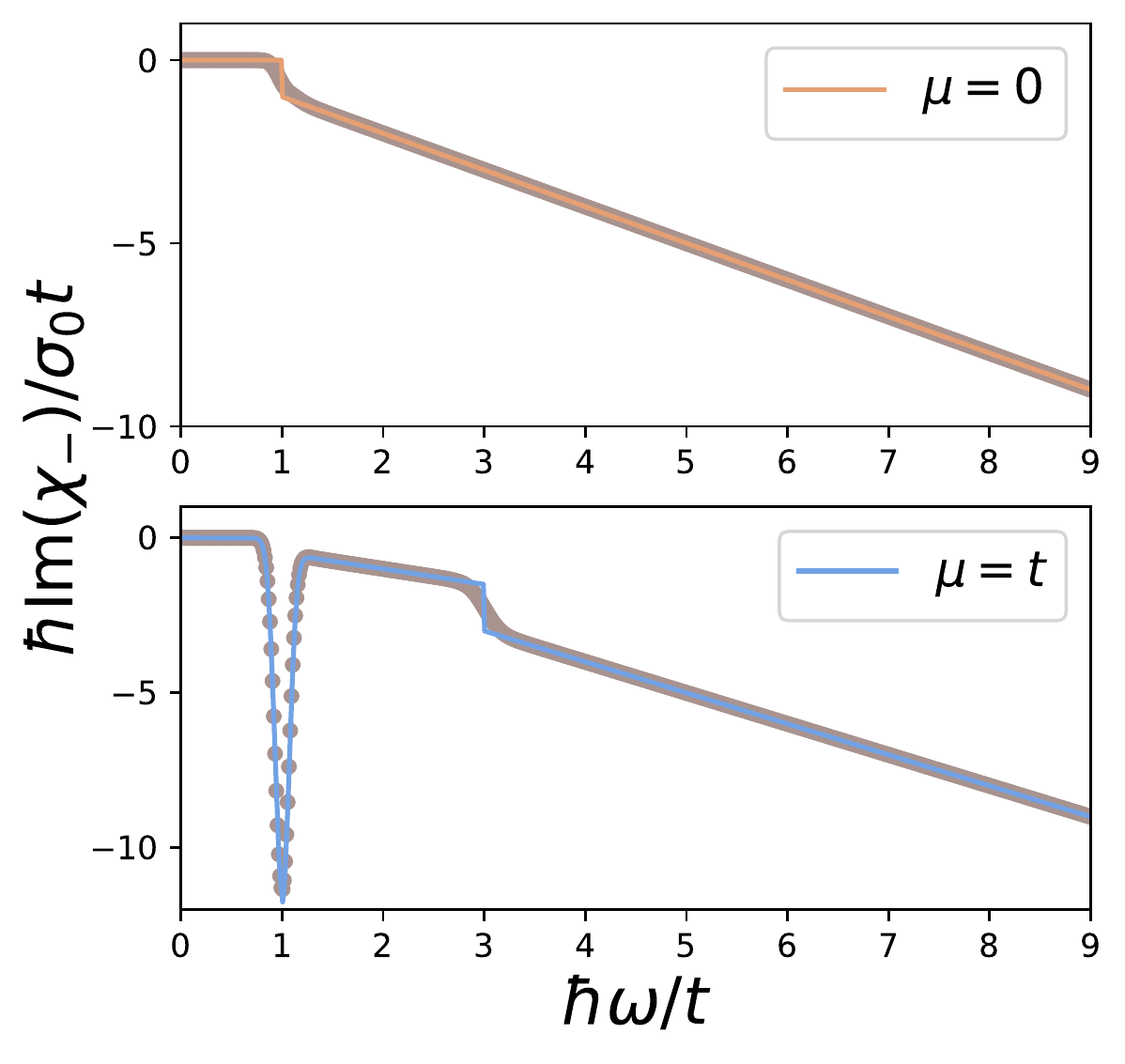}
    \includegraphics[width=.95\linewidth]{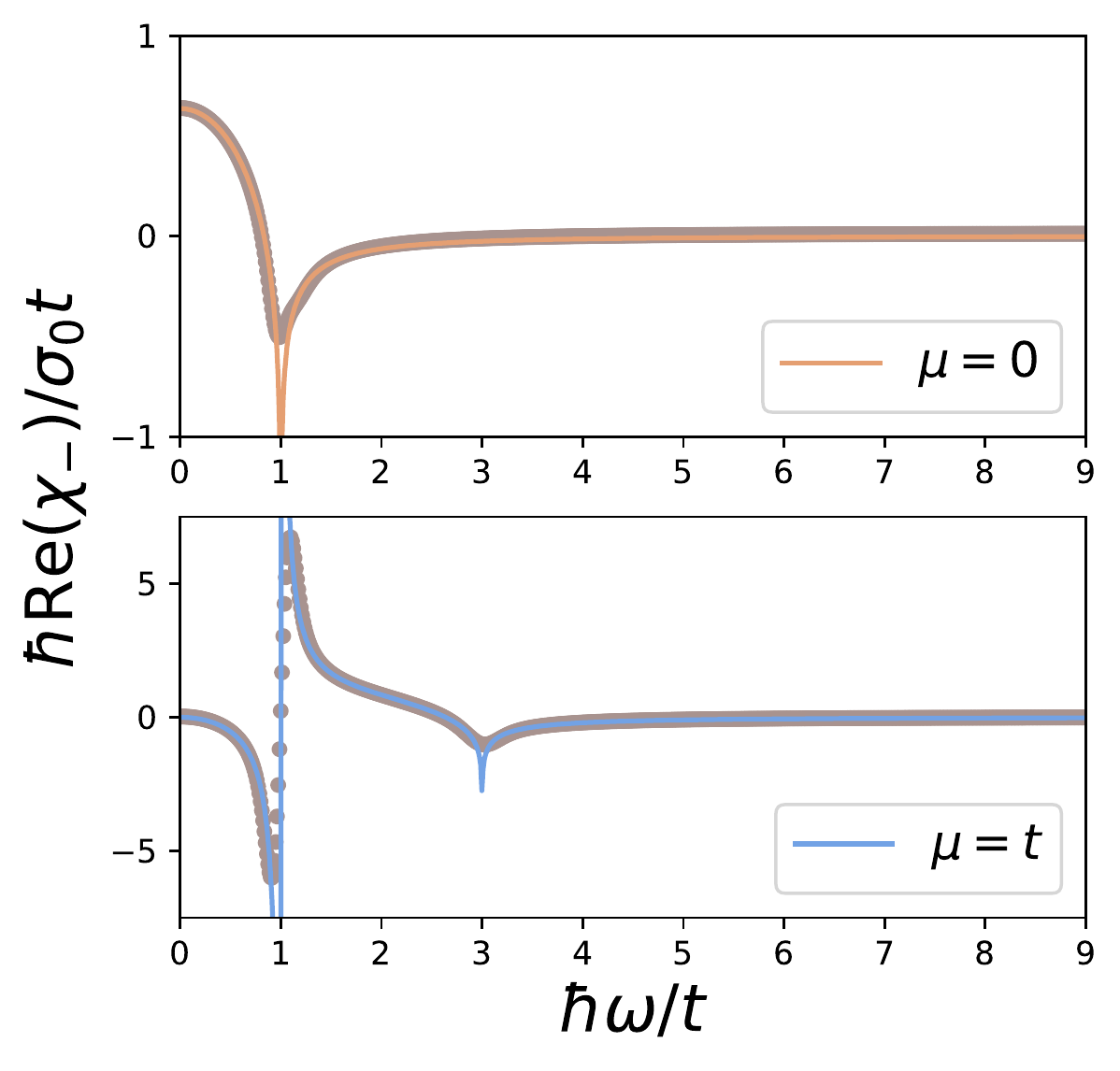}
    \caption{Same as in Fig. \ref{chi}, but for AB-stacked bilayer graphene.}
    \label{cond}
\end{figure}
With $U=t\left(
\begin{array}{cc}
0&0\\
1&0
\end{array}\right)$, we describe AB-stacked graphene whose optical conductivity has been discussed in Ref. \onlinecite{Abergel07} and the plasmonic properties in Ref. \onlinecite{Low14b}. 

The eigenenergies are now given by $E_{k,n_1,n_2}=\frac{1}{2}(n_1t+n_2\xi_k)$ with $n_1,n_2=\pm$ and $\xi_k=\sqrt{(2\hbar v_Fk)^2+t^2}$. The eigenvectors are
\begin{align}
\psi_\k^{+,\pm}&=N\left(\frac{t-\xi_k}{2\hbar v_F k},
    \mp e^{i\varphi_\k}, e^{-i\varphi_\k},\mp \frac{t-\xi_k}{2\hbar v_F k}\right)^T \nonumber \\
\psi_{\k}^{-,\pm} &= N\left( 1,\mp\frac{t-\xi_k}{2\hbar v_F k}e^{i\varphi_\k},  -\frac{t-\xi_k}{2\hbar v_F k}e^{-i\varphi_\k},\pm1\right)^T, \nonumber
\end{align}
where $\varphi_\k$ denotes the angle that $\k$ forms with the $x$-axis and $N = (2(1+(\frac{t-\xi_k}{2\hbar v_F k})^2))^{-\frac{1}{2}}$. The velocity operator of layer $\ell$ is again given by $\hat\v_\ell={\bm\sigma}\otimes{\bm1}_\ell$ and the matrix elements of the current counterflow read $v^-_{\k++,\k+-} = iv_F\sin(\varphi_\k)$, $v^-_{\k++,\k-+} = -v_F\cos(\varphi_\k)$, $v^-_{\k+-,\k--} = v_F \cos(\varphi_\k)$, $v^-_{\k+-,\k--} = iv_F \sin(\varphi_\k)$
and zero otherwise. Again, there is a perfect nesting condition, this time at $\hbar\omega=t$. This will also give rise to collective magnetic excitations and the resonant contribution of the magnetic response reads
\begin{align}
 \chi_{-}^{nested}(\omega) =&-\frac{e^2g_sg_v}{32\pi\hbar^2}\bigg[\frac{2t}{t^2-(\hbar\omega)^2}+i\pi\delta(\hbar\omega-t) \bigg] \\\nonumber  \times& \Big[2|\mu|(|\mu|+t)\theta(t-|\mu|)  +4|\mu|t\theta(|\mu|-t) \Big]\;.
\end{align}
However, contrary to the AA-stacked bilayer, perfect nesting also occurs in the optical absorption of AB-stacked graphene and allows the existence of high-energy electronic or charged plasmons with frequency $\omega \approx  t/\hbar $.\cite{Low14b}

In Fig. \ref{cond}, we show the real and imaginary part of the full magnetic response as discussed in Appendix A for $\mu=0$ and $\mu=t$. The delta-function of the imaginary part of the analytical solution (full line) is broadened by the same value ($\eta = 0.1t$) as used in the numerical solution (dots). The singularity in the real part of the analytical solution, however, is plotted for the clean system and for $\mu=0$, this singularity is logarithmic, whereas for $\mu=t$ it is algebraic.

\subsection{Numerical estimates}
As discussed above, the transverse antisymmetric photonic propagator is given by $d_{t-}=-(1-e^{-q'a})/2\epsilon_0c^2q'\approx -a/2\epsilon_0c^2$, and plasmonic excitations obey $1-d_{t-}\chi_- =0$. Therefore, only a negative real response, $\re\chi_{-}(\omega)<0$, allows for solutions. True dissipationless plasmons further demand $\im\chi_{-}(\omega)=0$, and a finite value of $\im\chi_{-}(\omega)$ yields damped plasmons with plasmon frequency $\omega_p$ given by $1-d_{t-}\re\chi_{-}(\omega_p)=0$, and inverse lifetime $\gamma =\im\chi_{-}(\omega_p)/\frac{\partial}{\partial\omega}  \re\chi_{-}(\omega_p)$.\cite{Stauber14b}

Setting the doping level to $\mu=t$, the real part of the magnetic current response near the nesting frequency reads
\begin{align}
    \re\chi_{-}^{AA}(\omega) &= \frac{16\sigma_0t}{\pi \hbar}\frac{1}{(\hbar \omega/t)^2-4}\;,\\
    \re\chi_{-}^{AB}(\omega) &= \frac{4\sigma_0t}{\pi \hbar}\frac{1}{(\hbar \omega/t)^2-1}\;,
\end{align}
where $\sigma_0=e^2/4\hbar$ is the universal conductivity of graphene.\cite{Nair08}

\begin{figure}
    \centering
    \includegraphics[width=.95\linewidth]{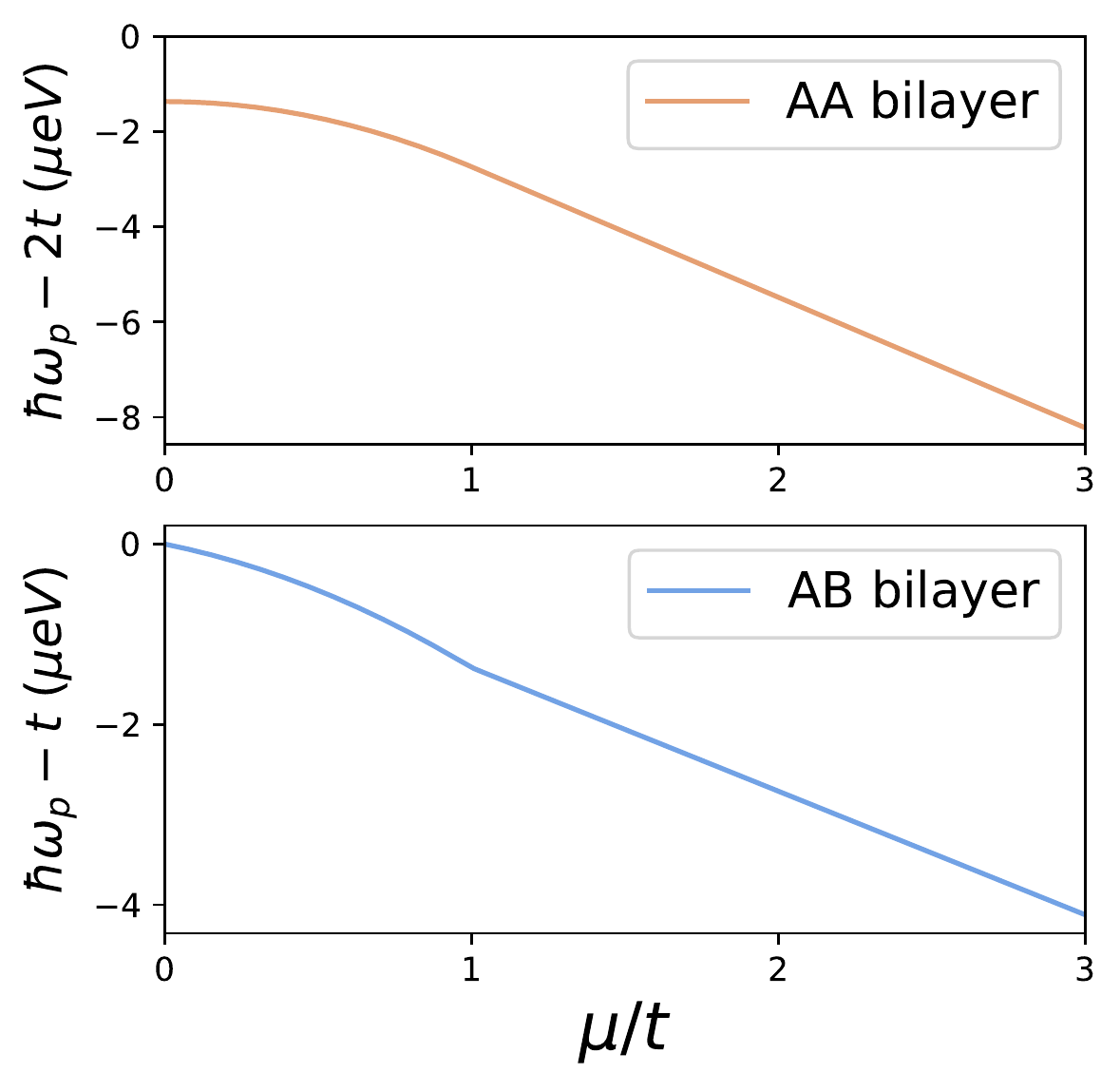}
    \caption{Frequency of the magnetic plasmons($\hbar \omega_p$) for $AA$ and $AB$ bilayer graphene as a function of the Fermi level. We set $t=0.33$ eV. In both cases, we plot the difference between the palsmon frequencies and the nesting frequency.}
    \label{omegaplasmon}
\end{figure}

With the fine-structure constant $\alpha=e^2/4\pi\epsilon_0\hbar c \sim 1/137$, $\hbar c=1973$eV\AA, $a=3.4$\AA${}$ and $t=0.33$eV, we thus have the following plasmonic resonances:
\begin{align}
\hbar\omega_p^{AA}&=\sqrt{1-\alpha\frac{2at}{\hbar  c}}2t = \sqrt{1-8.3\times10^{-6}}2t\;,\\
\hbar\omega_p^{AB}&=\sqrt{1-\alpha\frac{2at}{\hbar  c}}t = \sqrt{1-8.3\times10^{-6}}t\;.
\end{align}
The plasmon frequencies are thus separated from the resonant or nesting energy by  $2.7\mu \textrm{eV}$ and $1.4\mu \textrm{eV}$ for AA- and AB-stacked bilayers, respectively, which obviously calls for ultra-clean samples. In Fig. \ref{omegaplasmon}, we show its dependence on the chemical potential.

Since the broadening of the delta-function is of the order of the free-mean path, the pole is normally superposed by the dissipative delta-function in the imaginary part. So, even though the predicted modes are not likely to be observed in actual bilayer graphene samples, these excitations might be observable in artificial systems such as cold atoms. E.g., if we set the separation between the layers to be $a=100$nm, then we get $2t-\hbar\omega_p=0.8$meV ($0.4$meV) for AA (AB) bilayers which should be observable. 
\section{Magnetic instability in neutral AA-stacked bilayer graphene}
\label{MagneticInstability}
The antisymmetric response of AA-stacked bilayer graphene at the neutrality point $\mu=0$ is logarithmically diverging at $\omega=0$, see Eq. (\ref{responseAA}). This hints to a mode condensation or Condon instability as discussed in Refs. \onlinecite{Nataf19,Andolina20}. The relevant equilibrium response function requires the order of limit to be $\omega\to0$ and then $q\to0$, opposite to the one calculated here. The difference between the two limit orders is a contact term related to the density of states, a finite magnitude that cannot compete with the divergence.\cite{Stauber18b} Therefore, we can use our previous results for the ground-state stability analysis. 

Now we estimate the critical temperature. We start from the RPA instability $d_{t,-}\chi_-=1$, which can be thought of as a Stoner criterium. A proper analysis would require the calculation of a finite temperature response. On physical grounds, the latter can be inferred from the replacement $\hbar\omega\to k_BT$. For $\mu=0$ and $\omega\to0$, we have $\chi_-=t\frac{e^2}{\pi\hbar^2}\ln\frac{2t}{k_BT}$, see Appendix A. We thus obtain for the critical temperature
\begin{align}
k_BT_c=2t\exp\left(-\frac{3}{4\alpha}\frac{t_0}{t}\frac{a_0}{a}\frac{c}{v_F}\right)\;,
\end{align}
where $t_0\sim2.7$eV is the in-plane hopping amplitude and $a_0=1.42$\AA${}$ the carbon-carbon distance. 

The critical temperature is virtually zero, however, it may increase after renormalization of the single-particle parameters in twisted bilayer graphene where the electron density is concentrated at the regions of local AA-stacking. In this context, we mention that orbital ferromagnetism has recently been observed for filling fraction $n=3/4$ of the first conduction band of twisted bilayer graphene.\cite{Sharpe21} 

Our observation also links with a recent publication stating that the divergent paramagnetic response in twisted bilayer graphene can lead to a permanent in-plane magnetic moment.\cite{Guerci21} Note that this is not possible in one layer with a perpendicular magnetic moment.\cite{Andolina20} Also note that this instability is different from the one due to the short-ranged Hubbard interaction.\cite{Rakhmanov12,Brey13,Pena14} Last, due to perfect screening, this mode cannot be enhanced by a superconducting cavity,\cite{Nataf19} see Appendix C.
\section{Conclusions}
In this work, we have discussed the magnetic response of AA- and AB-stacked bilayer. For both stacking forms, we find a resonance due to perfect nesting which gives rise to an algebraic divergence in the real part of the magnetic instability. Within RPA, this gives rise to an oscillating in-plane magnetic moment that might be observable in ultra-clean samples. In AA-stacked samples, the ground-state is further given by a symmetry-broken state leading to in-plane orbital ferromagnetism. By identifying in the electronic structure of simple models the key features for appearance, as we do here, we thus offer physical guidance for its search in other systems.

The magnetic excitations/instabilities are composed of counterflow currents in each layer, giving rise to a magnetic moment parallel to the planes, thus providing a clear intuition of the magnetic nature in contrary to the transverse excitations in a single layer. Also, they do not depend on the external dielectric environment including optical cavities which should help to stabilize them, thus paving the way to novel plasmonics that is not limited by charged impurities. Last, these excitations might play a role in magic angle twisted bilayer samples as their electronic properties are mostly determined by confined AA-stacked regions.  

\section{Acknowledgments}
This work has been supported by Spain's MINECO under Grants No. FIS2017-82260-P,  PGC2018-096955-B-C42, PID2020-113164GB-I00, and CEX2018-000805-M as well as by the CSIC Research Platform on Quantum Technologies PTI-001. The access to computational resources of CESGA (Centro de Supercomputaci\'on de Galicia) is also gratefully acknowledged. 

\begin{widetext}
\appendix
\counterwithin{figure}{section}
\section{MAGNETIC RESPONSE IN AA AND AB-STACKED BILAYER GRAPHENE.}
\subsection{AA-stacked bilayer graphene}
In Fig. \ref{bands}, we represent and label the allowed transitions of the magnetic current operator for AA and AB-bilayers. In the $T=0$ limit, the Fermi distribution becomes the step function $\theta(\mu-\epsilon)$ and taking $\eta \rightarrow 0$ in Eq. \ref{kubo} allows for an easy evaluation of the imaginary part of the response function. The non resonant transitions yield the non-resonant or not nested part of the response function:
\begin{align}
    \im\chi_{-}^{not nested}(\omega) = -\pi\frac{e^2g_sg_v}{2A}\frac{A}{(2\pi)^2}(\pi v_F^2) \int k dk \Bigg\{\delta(\hbar \omega - (2t - 2\hbar v_Fk)) \theta\left(\frac{t-|\mu|}{\hbar v_F}-k\right) + \nonumber \\ \delta(\hbar \omega - (2t + 2\hbar v_Fk)) \theta\left(k - \frac{|\mu|-t}{\hbar v_F}\right) + \delta(\hbar \omega - (-2t + 2\hbar v_Fk)) \theta\left(k - \frac{|\mu|+t}{\hbar v_F}\right)\Bigg\}.
\end{align}
\begin{figure}[b]
    \begin{minipage}{.4\linewidth}
    \centering
    \includegraphics[width=1\linewidth]{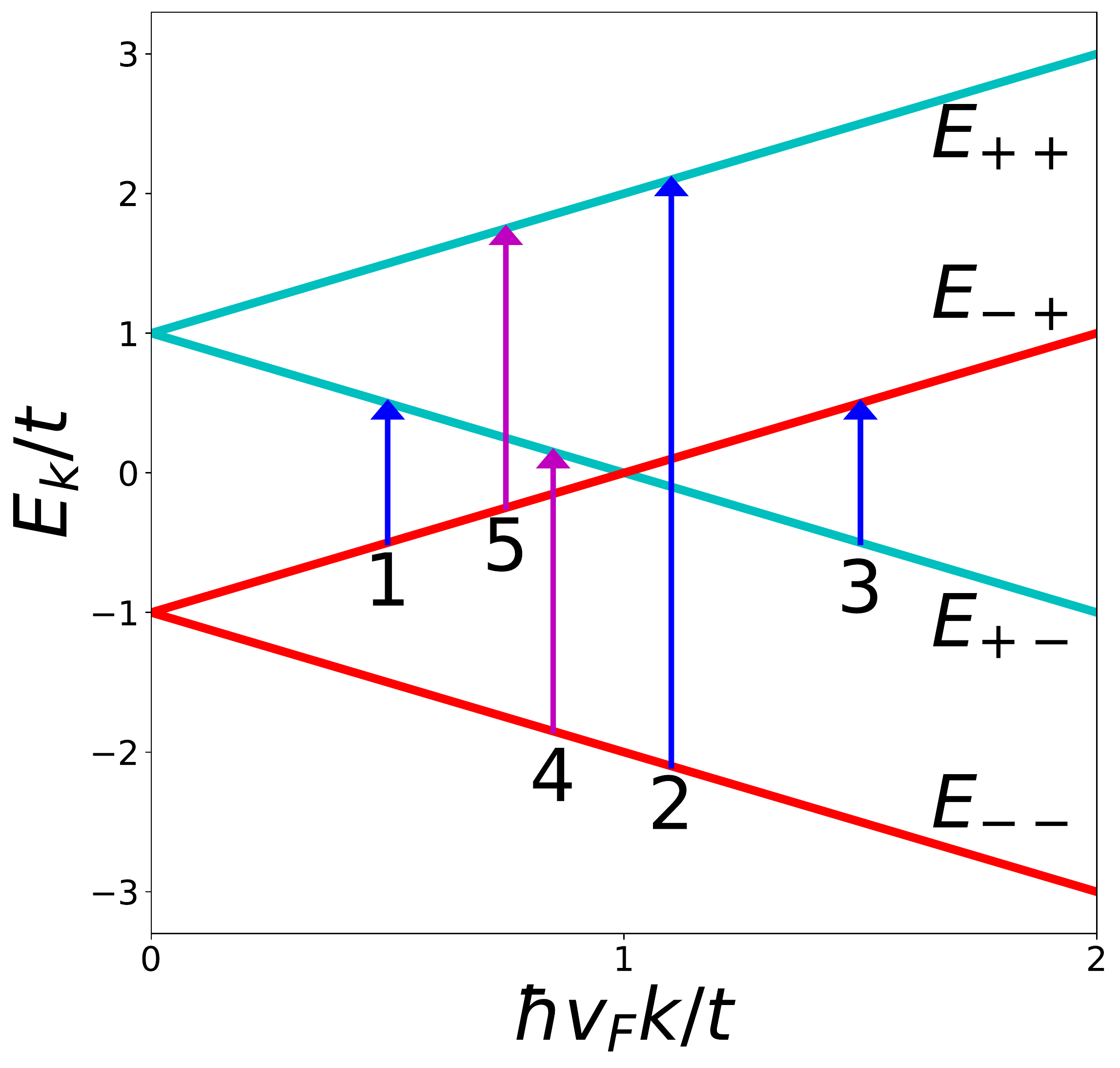}
    \end{minipage}
    \quad
    \begin{minipage}{.4\linewidth}
    \centering
    \includegraphics[width=1\linewidth]{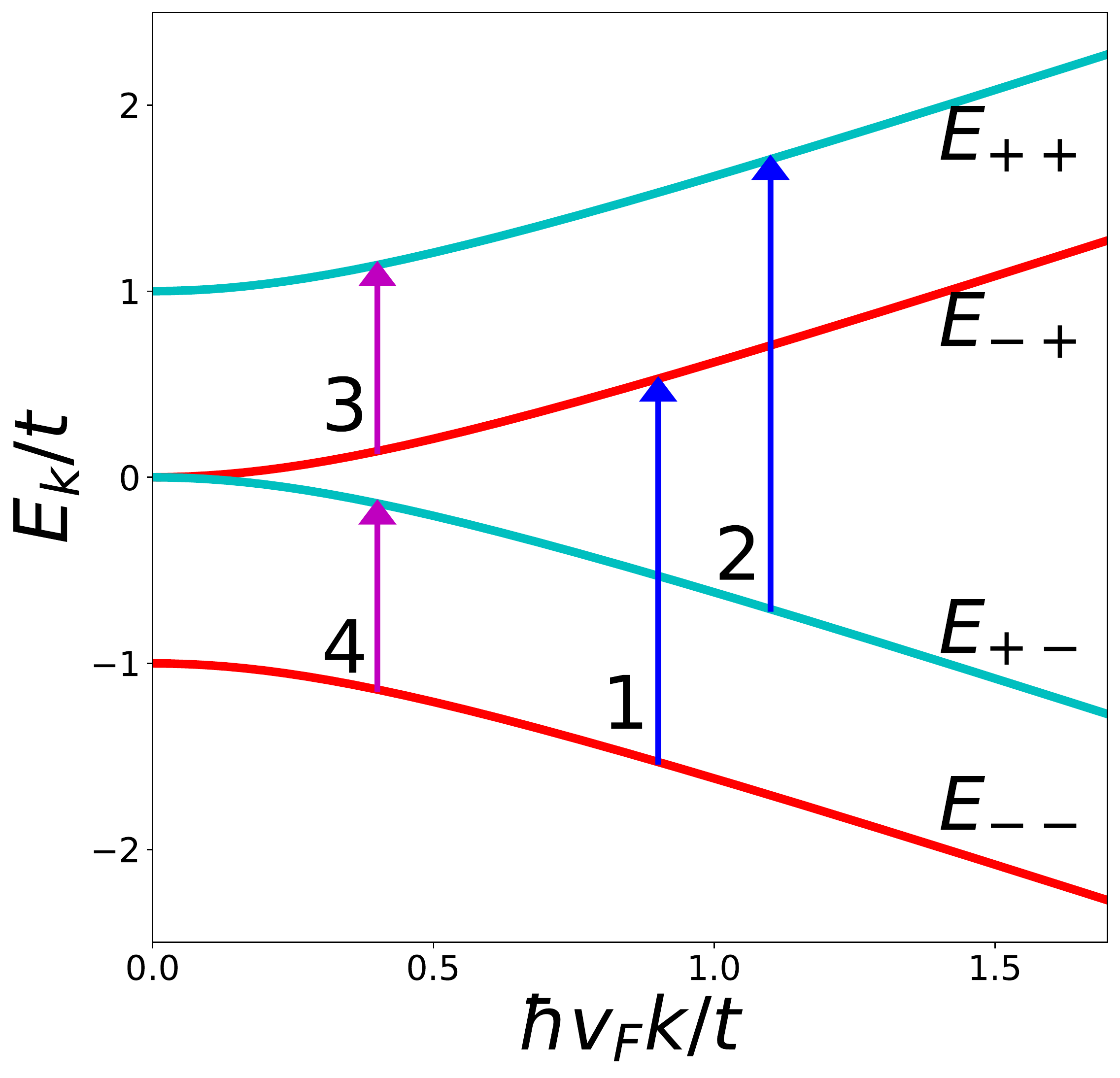}
    \end{minipage}
    \label{bands}
    \caption{Band structure of AA(left), and AB(right) stacked bilayer graphene, and allowed transitions of the current counterflow operator. Bonding states ($n_1=-1$) are depicted in red, and anti-bonding states ($n_1=+1$) in blue. Notice the perfect nesting of transitions 4 and 5 in AA, and 3 and 4 in AB-stacked graphene. 
    }
\end{figure}
In prefactor $\pi v_F^2$ comes from the angular integral of $\frac{v_F^2}{4}(2+2\cos(2\phi_k))$. Above, the first, second and third term are the contribution of transitions 1, 2 and 3 respectively. Performing the integrals gives
\begin{align}
    \im\chi_{-}^{not nested}(\omega) = -\frac{e^2v_F^2}{2}\Bigg\{&\frac{2t-\hbar \omega}{(2\hbar v_F)^2}\theta(\hbar \omega - 2|\mu|)\theta(2t - \hbar \omega) + \frac{\hbar \omega - 2t}{(2\hbar v_F)^2}\theta(\hbar \omega - 2|\mu|)\theta(\hbar \omega - 2t) + \nonumber \\ &\frac{2t+\hbar \omega}{(2\hbar v_F)^2}\theta(\hbar \omega - 2|\mu|)\theta(2t + \hbar \omega) \bigg\}.
\end{align}
\begin{figure}
    \begin{minipage}{.45\linewidth}
    \centering
    \includegraphics[width=1\linewidth]{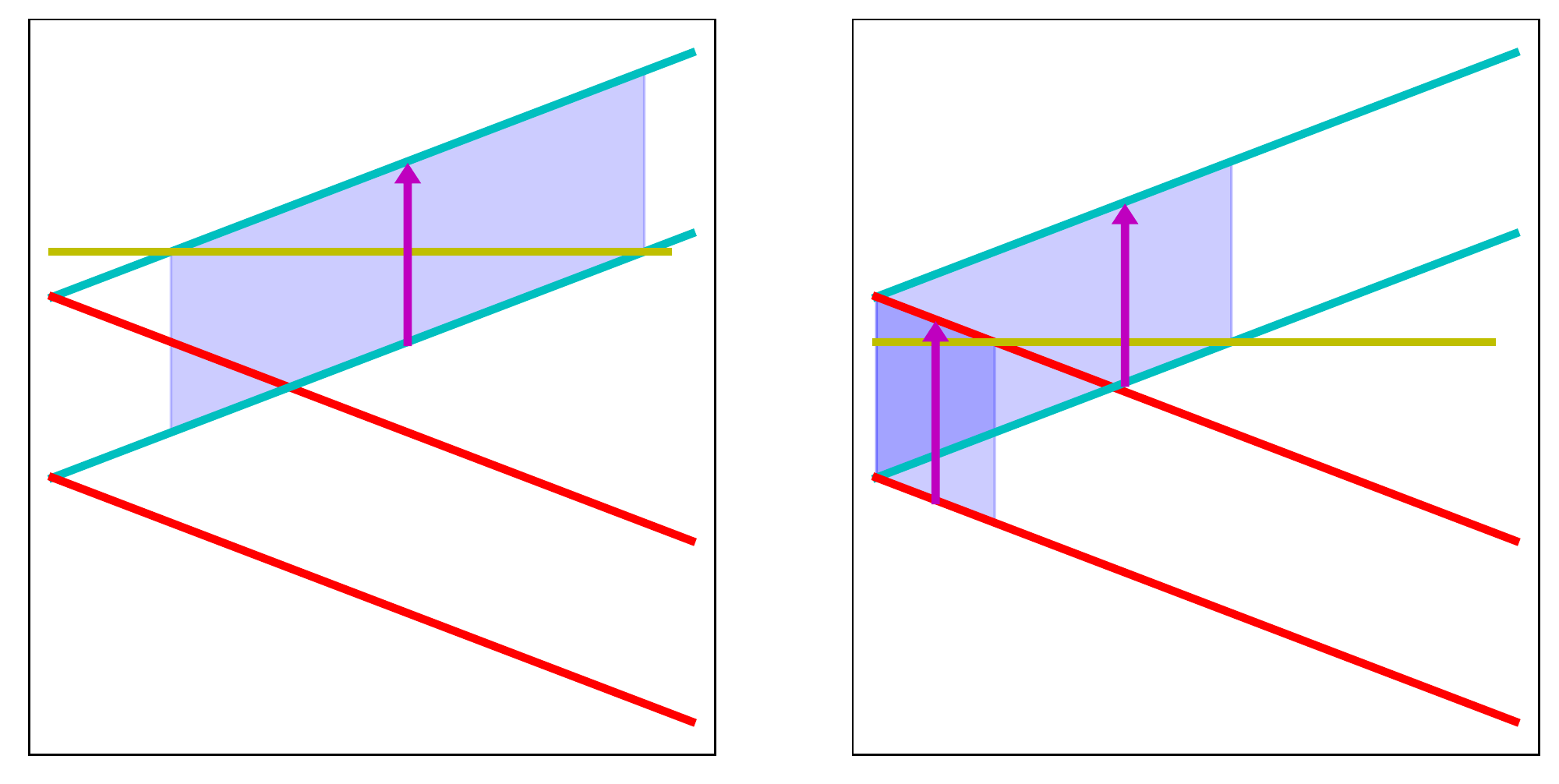}
    \end{minipage}
    \quad
    \begin{minipage}{.45\linewidth}
    \centering
    \includegraphics[width=1\linewidth]{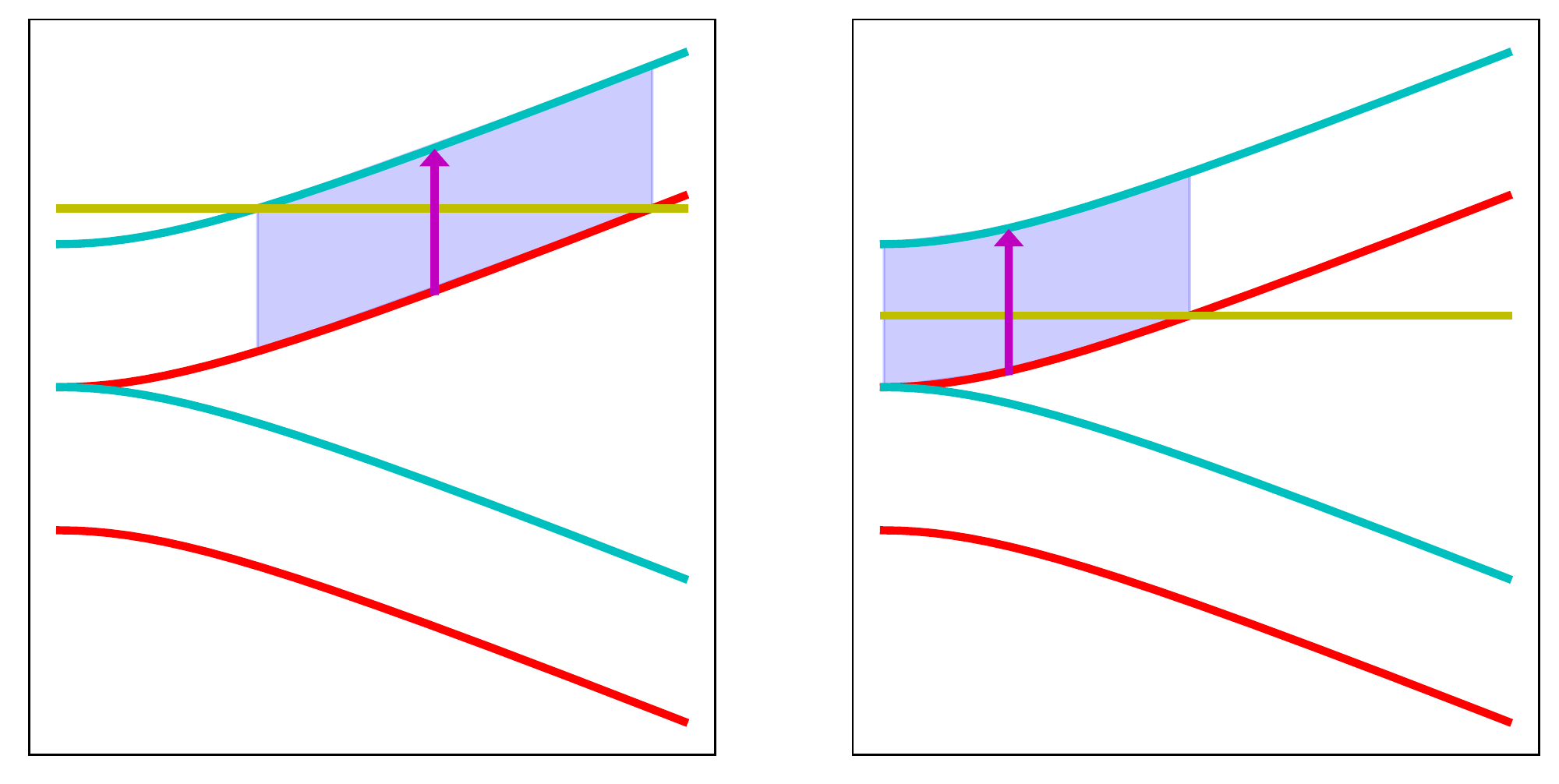}
    \end{minipage}
    \label{bands}
    \caption{Nested transitions that contribute to the resonance of the magnetic response in AA and AB bilayer graphene, for $\mu > t$ and $\mu < t$. Notice that at charge neutrality the weight of the resonance becomes zero for AB graphene, whereas it is always non-zero in AA bilayer.}
\end{figure}

For the resonant transitions, the imaginary part becomes:
\begin{align}
    \im\chi_{-}^{nested}(\omega) = -\pi\frac{e^2g_sg_v}{2A}\frac{A}{(2\pi)^2}(\pi v_F^2) \int k dk \Bigg\{&\delta(\hbar \omega - 2t) \theta\left(k - \frac{-(t+\mu)}{\hbar v_F}\right)\theta\left( \frac{t-\mu}{\hbar v_F} - k\right) + \nonumber \\ &\delta(\hbar \omega - 2t) \theta\left(k - \frac{\mu-t}{\hbar v_F}\right) \theta\left(\frac{\mu + t}{\hbar v_F} - k \right) \Bigg\}.
\end{align}

The first and second term are the contribution of transitions 4 and 5 respectively. Equivalently,
\begin{align}
    \im\chi_{-}^{nested}(\omega) =      -\frac{e^2g_sg_v}{8} \frac{v_F^2}{(2\hbar v_F)^2} \delta(\hbar \omega - 2t)\bigg\{4(\mu^2 + t^2)\theta(t-|\mu|) +8|\mu|t\theta(|\mu| - t) \bigg\}.
\end{align}

The total response is the sum of the non resonant and the resonant part:
\begin{align}
\begin{split}
     \im\chi_{-}(\omega) = -\frac{e^2g_sg_v}{16\hbar^2} \bigg\{ \Big[ \hbar \omega \theta(\hbar \omega - 2t) + 2t \theta(2t-\hbar \omega)  \Big] \theta(\hbar \omega - 2|\mu|) + \Big[ 2(\mu^2 + t^2) \theta(t-|\mu|) + 4t|\mu|\theta(|\mu| -t) \Big] \delta(\hbar \omega - 2t)\bigg\}.
     \end{split}
\end{align}

In terms of the dimensionless energy, $x=\frac{\hbar \omega}{t}$, and the universal conductivity of graphene, $\sigma_0 = \frac{e^2}{4\hbar}$,\cite{Nair08} yields with $g_s=g_v=2$
\begin{align}
    \im\chi_{-}(x) = -\frac{\sigma_0t}{\hbar} \left\{ x\left[\theta(x- 2) + \frac{2}{x} \theta(2-x)  \right] \theta(x - 2|\mu|/t) + \left[2\left(\frac{\mu^2}{t^2} + 1\right) \theta(1-|\mu|/t) + \frac{4
    |\mu|}{t} \theta(|\mu|/t -1) \right] \delta(x - 2)\right\}\;. 
\end{align}

For the real part we use the Kramers-Kronig relation 
\begin{equation}
    \re\chi_{-}(\omega) = \frac{2}{\pi}\mathcal{P}\int_0^\infty d\omega' \im\chi_{-}(\omega') \frac{\omega'}{\omega'^2 - \omega^2},
\end{equation}
where $\mathcal{P}$ denotes the Cauchy principal value.

The delta function can be integrated directly which gives the contribution of the resonant transitions to the real part presented in the main text,
\begin{align}
    \re\chi_{-}^{nested}(\omega) = \frac{-e^2g_sg_v}{8\pi \hbar^2} \frac{2t}{(2t)^2 - (\hbar\omega)^2}\Big[ 2(\mu^2 + t^2) \theta(t-|\mu|) + 4t|\mu|\theta(|\mu| -t) \Big]\;.
\end{align}

For the non resonant transitions, we perform the following integrals:
\begin{align}
    \re\chi_{-}^{not nested}(\omega) =& \frac{-e^2g_sg_v}{8\pi \hbar^2} \mathcal{P} \int_{2|\mu|/\hbar}^\infty d\omega' \frac{\omega'}{\omega'^2-\omega^2} \Big[ \hbar \omega' \theta(\hbar \omega' -2t) + 2t \theta(2t - \hbar \omega')\Big]  \nonumber \\
    =&\frac{-e^2g_sg_v}{8\pi \hbar^2}\Bigg\{\hbar \omega \mathcal{P}\int_{\max(2|\mu|/\hbar\omega,2t/\hbar\omega)}^\infty dy \frac{y^2}{y^2-1} + t \mathcal{P}\int_{(2|\mu|/\hbar\omega)^{2}}^{(2t/\hbar\omega)^{2}}  \frac{dz}{z-1}\theta(t-|\mu|) \Bigg\},
\end{align}
with $y=\omega'/\omega$ and $z=(\omega'/\omega)^2$. The first integral diverges in the Dirac cone approximation, so we proceed by separating the divergent part $y^2/(y^2-1) = 1 + 1/(y^2-1)$ and integrate up to a frequency cutoff $\Lambda$. 
With this scheme, the first term of the real part gives
\begin{align}
    \hbar \omega \mathcal{P}\int_{\max(2|\mu|/\hbar\omega,2t/\hbar\omega)}^\infty dx \frac{x^2}{x^2-1} = \hbar \Lambda -\max(2t,2|\mu|) -  \frac{\hbar \omega}{2}\log\bigg(\bigg|\frac{\hbar\omega - \max(2t, 2|\mu|)}{\hbar\omega+ \max(2t, 2|\mu|)}\bigg|\bigg),
\end{align}
and the second term can easily be integrated.
The total real part of $\chi_{-}$ is thus
\begin{align}
    \re\chi_{-}(\omega) = \frac{-e^2g_sg_v}{8\pi\hbar^2} \Bigg\{& - \max(2t,2|\mu|)- \frac{\hbar \omega}{2}\log\bigg(\bigg|\frac{\hbar\omega - \max(2t, 2|\mu|)}{\hbar\omega+ \max(2t, 2|\mu|)}\bigg|\bigg) + t \log\bigg(\bigg|\frac{(2t)^2-(\hbar\omega)^2}{(2|\mu|)^2-(\hbar\omega)^2}\bigg|\bigg) \theta(t-|\mu|) \nonumber \\& +  \frac{2t}{(2t)^2-(\hbar\omega)^2}\big[2(\mu^2 + t^2) \theta(t-|\mu|) + 4t|\mu|\theta(|\mu| -t) \big]\Bigg\}.\label{responseAA}
\end{align}

We have removed the part proportional to $\Lambda$ above, because the constant contribution of the response due to the diamagnetic currents cancels exactly the cut-off dependent term, as required by gauge invariance. \\
Again, with $g_s, g_v=2$ and in terms of $x=\frac{\hbar \omega}{t}$ and $\sigma_0=\frac{e^2}{4\hbar}$, the real part reads
\begin{align}
     \re\chi_{-}(x) = -\frac{\sigma_0t}{\pi\hbar} \bigg\{& -4\max(1,|\mu|/t) - x\log\bigg(\bigg|\frac{x -2 \max(1, |\mu|/t)}{x + 2\max(1,|\mu|/t)}\bigg|\bigg) + 2 \log\bigg(\bigg|\frac{1-(x/2)^2}{(|\mu|/t)^2-(x/2)^2}\bigg|\bigg) \theta(1-|\mu|/t) \nonumber \\& +  \frac{8}{4-x^2}\left[\left(1+\frac{\mu^2}{t^2}\right) \theta(1-|\mu|/t) + \frac{2|\mu|}{t}\theta(|\mu|/t -1) \right]\bigg\}.
\end{align}
\subsection{AB-stacked bilayer graphene}
Similarly to AA-stacked graphene, the imaginary part of the magnetic response can be decomposed into a sum of non resonant and resonant transitions. The non resonant or not-nested part (excitations 1 and 2 in Fig.\ref{bands}) reads
\begin{align}
    \im\chi_{-}^{not nested}(\omega) = -\pi \frac{e^2g_sg_v}{2A} \frac{A}{(2\pi)^2}(\pi e^2 v_F^2) \int k dk \{\delta(\hbar \omega - \xi_k)\theta(\xi_k-2|\mu+t/2|) + \delta(\hbar \omega - \xi_k)\theta(\xi_k-2|\mu-t/2|)\}.
\end{align}

The prefactor has the same origin as in the AA-stacked case. In this case, $\pi v_F^2$ is the angular integral of $v_F^2\cos^2(\phi_k)$ or $v_F^2\sin^2(\phi_k)$ depending on the transition. Performing the integral yields
\begin{align}
\im\chi_{-}^{not nested}(\omega) = -\frac{e^2g_sg_v}{8}v_F^2\bigg\{\frac{\hbar\omega}{(2\hbar v_F)^2}[\theta(\hbar \omega - 2|\mu+t/2|) + \theta(\hbar \omega - 2|\mu-t/2|)]\theta(\hbar \omega - t)\bigg\}.
\end{align}

On the other hand, the nested transistions 3 and 4 of Fig. \ref{bands} contribute with
\begin{align}
\begin{split}
    \im\chi_{-}^{nested}(\omega) 
    = -\frac{e^2g_sg_v}{8}v_F^2\delta(\hbar \omega - t)\int k dk\bigg\{
    \theta\left(\xi_k-(2\mu-t)\right)\theta\left(2\mu+t-\xi_k\right) + \theta\left(\xi_k-(-2\mu-t)\right)\theta\left(-2\mu+t-\xi_k\right) \bigg\}.
    \end{split}
    \end{align}

Keeping in mind that $\xi_k=\sqrt{t^2+(2\hbar v_F k)^2}$, the resonant part reads
\begin{align}
    \im\chi_{-}^{nested}(\omega) = \frac{-e^2g_sg_v}{32\hbar^2}\Bigg\{ \Big[2|\mu|(|\mu|+t)\theta(t-|\mu|)+4|\mu|t\theta(|\mu|-t) \Big]\delta(\hbar\omega-t)
    \Bigg\}.
\end{align}

Finally, the total imaginary part is the sum of both contributions which yields
\begin{align}
    \im\chi_{-}(\omega) = \frac{-e^2g_sg_v}{32\hbar^2}\Bigg\{&\hbar\omega\Big[\theta(\hbar \omega - 2|\mu+t/2|) + \theta(\hbar \omega - 2|\mu-t/2|)\Big]\theta(\hbar \omega - t)\nonumber \\ &+\Big[2|\mu|(|\mu|+t)\theta(t-|\mu|)+4|\mu|t\theta(|\mu|-t) \Big]\delta(\hbar\omega-t)
    \Bigg\}.
\end{align}

Using the dimensionless variable $x=\frac{\hbar \omega}{t}$ and the constant $\sigma_0 =\frac{e^2}{4\hbar}$, we get with $g_s=g_v=2$ 
\begin{align}
\begin{split}
    \im\chi_{-}(x) = -\frac{\sigma_0t}{2\hbar} \Bigg\{&x\Big[\theta(x - 2|\mu/t+1/2|) + \theta(x - 2|\mu/t-1/2|)\Big]\theta(x - 1)\\ &+\left[\frac{2|\mu|}{t}\left(\frac{|\mu|}{t}+1\right)\theta(1-|\mu|/t)+\frac{4|\mu|}{t}\theta(|\mu|/t-1) \right]\delta(x-1)
    \Bigg\}.
\end{split}
\end{align}

Again, we calculate the real part via the Kramers-Kronig relations. The delta function gives
\begin{align}
    \re\chi_{-}^{nested}(\omega)=-\frac{e^2g_sg_v}{16\pi\hbar^2}\bigg[2|\mu|(|\mu|+t)\theta(t-|\mu|)+4|\mu|t\theta(|\mu|-t) \bigg]\frac{t}{t^2-(\hbar\omega)^2}.
\end{align}

The non resonant part can be calculated in a way similar to the AA-stacked case and gives
\begin{align}
\re\chi_{-}^{not nested}(\omega)=-\frac{e^2g_sg_v}{16\pi\hbar^2}\bigg\{&2\hbar\Lambda - \Big(\max(t,2|\mu-t/2|)+\max(t,2|\mu+t/2|)\Big)+\frac{\hbar \omega}{2}\log\bigg(\bigg|\frac{\max(t,2|\mu-t/2|)-\hbar \omega}{\max(t,2|\mu-t/2|)+\hbar \omega}\bigg|\bigg)\nonumber \\&+\frac{\hbar \omega}{2}\log\bigg(\bigg|\frac{\max(t,2|\mu+t/2|)-\hbar \omega}{\max(t,2|\mu+t/2|)+\hbar \omega}\bigg|\bigg)\bigg\}.
\end{align}

After cancelling the cut-off term by the diamagnetic contribution, the total real part of the response function then reads
\begin{align}
\re\chi_{-}(\omega)=-\frac{e^2g_sg_v}{16\pi\hbar^2}\Bigg\{& - \left(\max(t,2|\mu-t/2|)+\max(t,2|\mu+t/2|)\right)+\frac{\hbar \omega}{2}\log\bigg(\bigg|\frac{\max(t,2|\mu-t/2|)-\hbar \omega}{\max(t,2|\mu-t/2|)+\hbar \omega}\bigg|\bigg)\nonumber \\&+\frac{\hbar \omega}{2}\log\bigg(\bigg|\frac{\max(t,2|\mu+t/2|)-\hbar \omega}{\max(t,2|\mu+t/2|)+\hbar \omega}\bigg|\bigg) + \bigg[2|\mu|(|\mu|+t)\theta(t-|\mu|)+4|\mu|t\theta(|\mu|-t) \bigg]\frac{t}{t^2-(\hbar\omega)^2} \Bigg\}.
\end{align}
In dimensionless variables this yields
\begin{align}
\re\chi_{-}(x)=-\frac{\sigma_0t}{\pi\hbar}\Bigg\{& - \Big(\max(1,2|\mu/t-1/2|)+\max(1,2|\mu/t+1/2|)\Big)+\frac{x}{2}\log\bigg(\bigg|\frac{\max(1,2|\mu/t-1/2|)-x}{\max(1,2|\mu/t-1/2|)+x}\bigg|\bigg)\nonumber \\&+\frac{x}{2}\log\bigg(\bigg|\frac{\max(1,2|\mu/t+1/2|)-x}{\max(1,2|\mu/t+1/2|)+x}\bigg|\bigg) + \bigg[\frac{2|\mu|}{t}\left(\frac{|\mu|}{t}+1\right)\theta(1-|\mu|/t)+\frac{4|\mu|}{t}\theta(|\mu|/t-1) \bigg]\frac{1}{1-x^2} \Bigg\}.
\end{align}
\section{PLASMONS IN INHOMOGENEOUS ENVIRONMENTS}
The transverse photonic Green's function when the bilayer lies between to dielectrics is given by\cite{Stauber12}
\begin{align}
{\bf \mathcal{D}_t} = -\frac{\mu_1\mu_2\mu_0q'}{N_t}\begin{pmatrix} \cosh(q'a) + \frac{q_2'}{\mu_2q'}\sinh(q'a) && 1 \\
1 &&  \cosh(q'a) + \frac{q_1'}{\mu_1q'}\sinh(q'a) \end{pmatrix},
\label{b1}
\end{align}
with $N_t = q'(\mu_2q_1'+\mu_1q_2')\cosh(q'a) + (q_1'q_2' + \mu_1\mu_2q'^2)\sinh(q'a)$, and $q_i'=\sqrt{q^2-\epsilon_i\mu_i(\omega/c)^2}$. In between the two graphene layers, we assume vacuum with wave number $q'=\sqrt{q^2-(\omega/c)^2}$. Plasmons are the solutions of det$({\bf 1}- {\bf \chi}{\bf \mathcal{D}_t}) =0$. The zero eigenvalues of ${\bf 1}- {\bf \chi}{\bf \mathcal{D}_t}$ are the symmetric and antisymmetric plasmonic excitations. Let us calculate those. Expanding in powers of the $q_i'a$'s, the Green`s function becomes
\begin{align}
    {\bf \mathcal{D}_t} =& -\frac{\mu_1\mu_2\mu_0a}{\mu_1q_2'a+\mu_2q_1'a}\left( 1-\frac{(q_1'a)(q_2'a)}{\mu_1q_2'a+\mu_2q_1'a}-\frac{\mu_1\mu_2(q'a)^2}{\mu_1q_2'a+\mu_2q_1'a}\right)\begin{pmatrix} 1+ \frac{q_2'a}{\mu_2}+\frac{(q'a)^2}{2} && 1 \\ 1 && 1+\frac{q_2'a}{\mu_1}+\frac{(q'a)^2}{2}
    \end{pmatrix}  .
    \label{Dt}
\end{align}

Hence, the eigenvalues ($\lambda'$) of the matrix
\begin{align}
\left(\begin{array}{cc}
\chi_{11} & \chi_{12} \\ \chi_{12} & \chi_{11}
\end{array}\right)\left(\begin{array}{cc}
1+ q_2'a/\mu_2+(q'a)^2/2 & 1 \\ 1 & 1+q_2'a/\mu_1+(q'a)^2/2
\end{array}\right)
\end{align}
are related to the eigenvalues($\lambda $) of ${\bf 1 - \chi \mathcal{D}_t}$ by
\begin{align}
\lambda =1 + \frac{\mu_1\mu_2\mu_0a}{\mu_1q_2'a+\mu_2q_1'a}\left( 1-\frac{(q_1'a)(q_2'a)}{\mu_1q_2'a+\mu_2q_1'a}-\frac{\mu_1\mu_2(q'a)^2}{\mu_1q_2'a+\mu_2q_1'a}\right) \lambda'.
\end{align}

Up to order $(q_i'a)^2$, these solutions read:
\begin{align}
    \lambda' = &(\chi_{11} + \chi_{12})  + \chi_{11}\left( \frac{(q'a)^2}{2} + \frac{1}{2}\left(\frac{q_1'a}{\mu_1} + \frac{q_2'a}{\mu_2}\right) \right) \nonumber \\\pm& \Bigg\{(\chi_{11} + \chi_{12}) + \chi_{12}\left( \frac{(q'a)^2}{2} + \frac{1}{2}\left(\frac{q_1'a}{\mu_1} + \frac{q_2'a}{\mu_2}\right)\right)  + \frac{\chi_{11} - \chi_{12}}{8}\left( \frac{q_1'a}{\mu_1} - \frac{q_2'a}{\mu_2} \right)^2\Bigg\}.
    \label{b4}
\end{align}

Therfore, the plasmonic resonances of the system are the zeroes of the following eigenvalues:
\begin{align}
     \lambda^- =  1+ \frac{\mu_0 a(\chi_{11} - \chi_{12})}{2}\left( 1- \frac{1}{4}\left(\frac{q_2'a}{\mu_2} + \frac{q_1'a}{\mu_1} \right)  + \mathcal{O}((q_i'a)^2) \right)    
     \label{b6} \hspace{1.8cm}
\end{align}
\begin{align}
    \lambda^+ =  1 + \frac{\mu_0 a}{q_1'a/\mu_1 + q_2'a/\mu_2}  
    \Bigg\{& 
    (\chi_{11} + \chi_{12})\Bigg(
    2 + \Bigg( \frac{\left(q_1'a/\mu_1 - q_2'a/\mu_2\right)^2 -  (2q'a)^2}{2\left(q_1'a/\mu_1 + q_2'a/\mu_2\right)}\Bigg) + 2 \Bigg(\frac{(q_1'a/\mu_1)(q_2'a/\mu_2) + (q'a)^2}{q'1_a/\mu_2 + q'_2a/\mu_2}\Bigg)^2 - \nonumber \\ &\Bigg(\frac{2(q'a)^2 + (q_1'a/\mu_1)(q_2'a/\mu_2)}{2} \Bigg) 
    + \frac{\chi_{11} -  \chi_{12}}{8}\left(q_1'a/\mu_1 - q_2'a/\mu_2\right)^2 \Bigg\}
    + \mathcal{O}((q_i'a)^2)
\end{align}

One thing to note is that this approximation breaks down if the dielectric media 1 and 2 are the same for $q_1' = 0$.\\ In that case, when approaching the slower light cone, $q_1'a$ and  $q_2'a$ tend to 0 while $q'a$ tends to a constant, and the terms with $(q'a)^2/(q_1'a/\mu_1 + q_2'a/\mu_2)$ are no longer small, in fact become infinite. However, the divergence kicks off at very small values of $q_1'a$ and we can trust the formula for essentially the entire dynamic range of $q$, as we will see.

A simple calculation tells us that $(q'a)^2/(2q_1'a/\mu_1) \sim 1$ when $qa \sim \omega a/c_1 + \mu_1^2/8 \big((c/c_1)^2-1\big)^2(c_1/c)(\omega a/c)^3$, or equivalently, $q_1'a \sim \mu_1/2\big((c/c_1)^2-1\big)\left(\omega a/ c\right)^2$(where $c_1$ is the speed of light in medium 1). Taking the frequency to be of order the interlayer coupling, $\hbar \omega = 0.5$ eV, we have $\omega a/c \sim 10^{-3}$, $qa \sim \omega a/c_1 + 10^{-9}$ and $q_1'a \sim 10^{-6}$. On the other hand, setting the momentum scale to, say $q_0 = G/100 \sim 3\times10^{8} \textrm{m}^{-1}$, where G is the modulus of the reciprocal lattice vector of graphene, and the interlayer distance to $a=3.4\textup{\r{A}}$ we have that our approximation is valid up to $q_1'/q_0 \sim 10^{-5}$, or $q/q_0 \sim 10^{-8}$ to the right of the light cone of the dielectric.\\
Setting $q_1'a, q_2'a=0$ in Equation \ref{b1}, the eigenvalues at the dielectric light cone are $\lambda^+ = 1 + 2\mu_0 a (\chi_{11}+\chi_{12})/(q'a)^2$ and $\lambda^- = 1- \mu_0 a(\chi_{11}-\chi_{12})/2$. In conclusion, the symmetric eigenvalue remains bounded and we can trust Equation \ref{b6} even at $q_1'=0$. 

On the other hand, if the dielectric media are not the same, the quantity $(q'a)^2/(q_1'a/\mu_1) + (q_2'a/\mu_2))$ will not blow up, and for refractive indices $|n_1 - n_2| \gtrsim 10^{-6}$ the approximation is good all the way up to $q_1' = 0$(being 1 the slower medium).

\section{INFLUENCE OF AN OPTICAL CAVITY.}
Let us consider the bilayer placed inside an optical cavity, with boundaries at $z = \pm L/2$. In such a setting, the Green's function or photonic propagator for the transverse fields of a source located a $z'$ takes the form
\begin{align}
    D_l(z,z') = \frac{\mu_0}{q'}\frac{\sinh(q'(z^>-L/2))\sinh(q'(z^<+L/2))}{\sinh(q'L)}\;,
\end{align}
for $-L/2 < z < L/2$ and $D_l(z,z') = 0$ otherwise (where $z^< = \min(z,z')$ and $z^> = \max(z,z')$). If the bilayer is located at $z=\pm a/2$, the Green's function of the double layer structure is
\begin{align}
    {\bf \mathcal{D}_t} = \frac{\mu_0}{q'\sinh(q'L)}\begin{pmatrix} \sinh((a-L)/2)\sinh((a+L)/2) && -\sinh^2((L-a)/2) \\
    -\sinh^2((L-a)/2) && \sinh((a-L)/2)\sinh((a+L)/2) \end{pmatrix}\;,
\end{align}
Transverse plasmons are obtained after the in-phase and out-of-phase (counterflow) combinations:
\begin{align}
    1 + \frac{\mu_0a \chi_+}{q'a} \frac{\cosh(a/2)\sinh((L-a)/2)}{\cosh(L/2)} = 0 \\
    1 + \frac{\mu_0a\chi_-}{q'a}  \frac{\sinh(a/2)\sinh((L-a)/2)}{\sinh(L/2)} = 0
\end{align}
The form factors are increasing functions of $L$, so the light-matter coupling is larger when there is no cavity. Hence, the presence of the cavity does not favour the presence of these elusive excitations, as was already pointed out in Ref. \onlinecite{Nataf19}. This is due to the "perfect" screening of the magnetic excitations. 
\end{widetext}

\begin{thebibliography}{48}%
\makeatletter
\providecommand \@ifxundefined [1]{%
 \@ifx{#1\undefined}
}%
\providecommand \@ifnum [1]{%
 \ifnum #1\expandafter \@firstoftwo
 \else \expandafter \@secondoftwo
 \fi
}%
\providecommand \@ifx [1]{%
 \ifx #1\expandafter \@firstoftwo
 \else \expandafter \@secondoftwo
 \fi
}%
\providecommand \natexlab [1]{#1}%
\providecommand \enquote  [1]{``#1''}%
\providecommand \bibnamefont  [1]{#1}%
\providecommand \bibfnamefont [1]{#1}%
\providecommand \citenamefont [1]{#1}%
\providecommand \href@noop [0]{\@secondoftwo}%
\providecommand \href [0]{\begingroup \@sanitize@url \@href}%
\providecommand \@href[1]{\@@startlink{#1}\@@href}%
\providecommand \@@href[1]{\endgroup#1\@@endlink}%
\providecommand \@sanitize@url [0]{\catcode `\\12\catcode `\$12\catcode
  `\&12\catcode `\#12\catcode `\^12\catcode `\_12\catcode `\%12\relax}%
\providecommand \@@startlink[1]{}%
\providecommand \@@endlink[0]{}%
\providecommand \url  [0]{\begingroup\@sanitize@url \@url }%
\providecommand \@url [1]{\endgroup\@href {#1}{\urlprefix }}%
\providecommand \urlprefix  [0]{URL }%
\providecommand \Eprint [0]{\href }%
\providecommand \doibase [0]{http://dx.doi.org/}%
\providecommand \selectlanguage [0]{\@gobble}%
\providecommand \bibinfo  [0]{\@secondoftwo}%
\providecommand \bibfield  [0]{\@secondoftwo}%
\providecommand \translation [1]{[#1]}%
\providecommand \BibitemOpen [0]{}%
\providecommand \bibitemStop [0]{}%
\providecommand \bibitemNoStop [0]{.\EOS\space}%
\providecommand \EOS [0]{\spacefactor3000\relax}%
\providecommand \BibitemShut  [1]{\csname bibitem#1\endcsname}%
\let\auto@bib@innerbib\@empty
\bibitem [{\citenamefont {Chen}\ \emph {et~al.}(2012)\citenamefont {Chen},
  \citenamefont {Badioli}, \citenamefont {Alonso-Gonzalez}, \citenamefont
  {Thongrattanasiri}, \citenamefont {Huth}, \citenamefont {Osmond},
  \citenamefont {Spasenovic}, \citenamefont {Centeno}, \citenamefont
  {Pesquera}, \citenamefont {Godignon}, \citenamefont {Zurutuza~Elorza},
  \citenamefont {Camara}, \citenamefont {de~Abajo}, \citenamefont
  {Hillenbrand},\ and\ \citenamefont {Koppens}}]{Chen12}%
  \BibitemOpen
  \bibfield  {author} {\bibinfo {author} {\bibfnamefont {J.}~\bibnamefont
  {Chen}}, \bibinfo {author} {\bibfnamefont {M.}~\bibnamefont {Badioli}},
  \bibinfo {author} {\bibfnamefont {P.}~\bibnamefont {Alonso-Gonzalez}},
  \bibinfo {author} {\bibfnamefont {S.}~\bibnamefont {Thongrattanasiri}},
  \bibinfo {author} {\bibfnamefont {F.}~\bibnamefont {Huth}}, \bibinfo {author}
  {\bibfnamefont {J.}~\bibnamefont {Osmond}}, \bibinfo {author} {\bibfnamefont
  {M.}~\bibnamefont {Spasenovic}}, \bibinfo {author} {\bibfnamefont
  {A.}~\bibnamefont {Centeno}}, \bibinfo {author} {\bibfnamefont
  {A.}~\bibnamefont {Pesquera}}, \bibinfo {author} {\bibfnamefont
  {P.}~\bibnamefont {Godignon}}, \bibinfo {author} {\bibfnamefont
  {A.}~\bibnamefont {Zurutuza~Elorza}}, \bibinfo {author} {\bibfnamefont
  {N.}~\bibnamefont {Camara}}, \bibinfo {author} {\bibfnamefont {F.~J.~G.}\
  \bibnamefont {de~Abajo}}, \bibinfo {author} {\bibfnamefont {R.}~\bibnamefont
  {Hillenbrand}}, \ and\ \bibinfo {author} {\bibfnamefont {F.~H.~L.}\
  \bibnamefont {Koppens}},\ }\href {\doibase 10.1038/nature11254} {\bibfield
  {journal} {\bibinfo  {journal} {Nature}\ }\textbf {\bibinfo {volume} {487}},\
  \bibinfo {pages} {77} (\bibinfo {year} {2012})}\BibitemShut {NoStop}%
\bibitem [{\citenamefont {Fei}\ \emph {et~al.}(2012)\citenamefont {Fei},
  \citenamefont {Rodin}, \citenamefont {Andreev}, \citenamefont {Bao},
  \citenamefont {McLeod}, \citenamefont {Wagner}, \citenamefont {Zhang},
  \citenamefont {Zhao}, \citenamefont {Thiemens}, \citenamefont {Dominguez},
  \citenamefont {Fogler}, \citenamefont {Neto}, \citenamefont {Lau},
  \citenamefont {Keilmann},\ and\ \citenamefont {Basov}}]{Fei12}%
  \BibitemOpen
  \bibfield  {author} {\bibinfo {author} {\bibfnamefont {Z.}~\bibnamefont
  {Fei}}, \bibinfo {author} {\bibfnamefont {A.~S.}\ \bibnamefont {Rodin}},
  \bibinfo {author} {\bibfnamefont {G.~O.}\ \bibnamefont {Andreev}}, \bibinfo
  {author} {\bibfnamefont {W.}~\bibnamefont {Bao}}, \bibinfo {author}
  {\bibfnamefont {A.~S.}\ \bibnamefont {McLeod}}, \bibinfo {author}
  {\bibfnamefont {M.}~\bibnamefont {Wagner}}, \bibinfo {author} {\bibfnamefont
  {L.~M.}\ \bibnamefont {Zhang}}, \bibinfo {author} {\bibfnamefont
  {Z.}~\bibnamefont {Zhao}}, \bibinfo {author} {\bibfnamefont {M.}~\bibnamefont
  {Thiemens}}, \bibinfo {author} {\bibfnamefont {G.}~\bibnamefont {Dominguez}},
  \bibinfo {author} {\bibfnamefont {M.~M.}\ \bibnamefont {Fogler}}, \bibinfo
  {author} {\bibfnamefont {A.~H.~C.}\ \bibnamefont {Neto}}, \bibinfo {author}
  {\bibfnamefont {C.~N.}\ \bibnamefont {Lau}}, \bibinfo {author} {\bibfnamefont
  {F.}~\bibnamefont {Keilmann}}, \ and\ \bibinfo {author} {\bibfnamefont
  {D.~N.}\ \bibnamefont {Basov}},\ }\href {\doibase 10.1038/nature11253}
  {\bibfield  {journal} {\bibinfo  {journal} {Nature}\ }\textbf {\bibinfo
  {volume} {487}},\ \bibinfo {pages} {82} (\bibinfo {year} {2012})}\BibitemShut
  {NoStop}%
\bibitem [{\citenamefont {Grigorenko}\ \emph {et~al.}(2012)\citenamefont
  {Grigorenko}, \citenamefont {Polini},\ and\ \citenamefont
  {Novoselov}}]{Grigorenko12}%
  \BibitemOpen
  \bibfield  {author} {\bibinfo {author} {\bibfnamefont {A.~N.}\ \bibnamefont
  {Grigorenko}}, \bibinfo {author} {\bibfnamefont {M.}~\bibnamefont {Polini}},
  \ and\ \bibinfo {author} {\bibfnamefont {K.~S.}\ \bibnamefont {Novoselov}},\
  }\href@noop {} {\bibfield  {journal} {\bibinfo  {journal} {Nature Photon.}\
  }\textbf {\bibinfo {volume} {6}},\ \bibinfo {pages} {749} (\bibinfo {year}
  {2012})}\BibitemShut {NoStop}%
\bibitem [{\citenamefont {Bludov}\ \emph {et~al.}(2013)\citenamefont {Bludov},
  \citenamefont {Ferreira}, \citenamefont {Peres},\ and\ \citenamefont
  {Vasilevskiy}}]{Bludov13}%
  \BibitemOpen
  \bibfield  {author} {\bibinfo {author} {\bibfnamefont {Y.~V.}\ \bibnamefont
  {Bludov}}, \bibinfo {author} {\bibfnamefont {A.}~\bibnamefont {Ferreira}},
  \bibinfo {author} {\bibfnamefont {N.~M.~R.}\ \bibnamefont {Peres}}, \ and\
  \bibinfo {author} {\bibfnamefont {M.~I.}\ \bibnamefont {Vasilevskiy}},\
  }\href {\doibase 10.1142/S0217979213410014} {\bibfield  {journal} {\bibinfo
  {journal} {Int. J. Mod. Phys. B}\ }\textbf {\bibinfo {volume} {27}},\
  \bibinfo {pages} {1341001} (\bibinfo {year} {2013})}\BibitemShut {NoStop}%
\bibitem [{\citenamefont {Stauber}(2014{\natexlab{a}})}]{Stauber14}%
  \BibitemOpen
  \bibfield  {author} {\bibinfo {author} {\bibfnamefont {T.}~\bibnamefont
  {Stauber}},\ }\href {\doibase 10.1088/0953-8984/26/12/123201} {\bibfield
  {journal} {\bibinfo  {journal} {Journal of Physics: Condensed Matter}\
  }\textbf {\bibinfo {volume} {26}},\ \bibinfo {pages} {123201} (\bibinfo
  {year} {2014}{\natexlab{a}})}\BibitemShut {NoStop}%
\bibitem [{\citenamefont {Gon\c{c}alves}\ and\ \citenamefont
  {Peres}(2016)}]{Peres16}%
  \BibitemOpen
  \bibfield  {author} {\bibinfo {author} {\bibfnamefont {P.~A.~D.}\
  \bibnamefont {Gon\c{c}alves}}\ and\ \bibinfo {author} {\bibfnamefont
  {N.~M.~R.}\ \bibnamefont {Peres}},\ }\href@noop {} {\emph {\bibinfo {title}
  {An Introduction to Graphene Plasmonics}}}\ (\bibinfo  {publisher} {World
  Scientific},\ \bibinfo {address} {Singapore},\ \bibinfo {year}
  {2016})\BibitemShut {NoStop}%
\bibitem [{\citenamefont {Low}\ \emph {et~al.}(2017)\citenamefont {Low},
  \citenamefont {Chaves}, \citenamefont {Caldwell}, \citenamefont {Kumar},
  \citenamefont {Fang}, \citenamefont {Avouris}, \citenamefont {Heinz},
  \citenamefont {Guinea}, \citenamefont {Martin-Moreno},\ and\ \citenamefont
  {Koppens}}]{Low17}%
  \BibitemOpen
  \bibfield  {author} {\bibinfo {author} {\bibfnamefont {T.}~\bibnamefont
  {Low}}, \bibinfo {author} {\bibfnamefont {A.}~\bibnamefont {Chaves}},
  \bibinfo {author} {\bibfnamefont {J.~D.}\ \bibnamefont {Caldwell}}, \bibinfo
  {author} {\bibfnamefont {A.}~\bibnamefont {Kumar}}, \bibinfo {author}
  {\bibfnamefont {N.~X.}\ \bibnamefont {Fang}}, \bibinfo {author}
  {\bibfnamefont {P.}~\bibnamefont {Avouris}}, \bibinfo {author} {\bibfnamefont
  {T.~F.}\ \bibnamefont {Heinz}}, \bibinfo {author} {\bibfnamefont
  {F.}~\bibnamefont {Guinea}}, \bibinfo {author} {\bibfnamefont
  {L.}~\bibnamefont {Martin-Moreno}}, \ and\ \bibinfo {author} {\bibfnamefont
  {F.}~\bibnamefont {Koppens}},\ }\href {http://dx.doi.org/10.1038/nmat4792}
  {\bibfield  {journal} {\bibinfo  {journal} {Nat Mater}\ }\textbf {\bibinfo
  {volume} {16}},\ \bibinfo {pages} {182} (\bibinfo {year} {2017})}\BibitemShut
  {NoStop}%
\bibitem [{\citenamefont {Koppens}\ \emph {et~al.}(2011)\citenamefont
  {Koppens}, \citenamefont {Chang},\ and\ \citenamefont {Garcia~de
  Abajo}}]{Koppens11}%
  \BibitemOpen
  \bibfield  {author} {\bibinfo {author} {\bibfnamefont {F.~H.~L.}\
  \bibnamefont {Koppens}}, \bibinfo {author} {\bibfnamefont {D.~E.}\
  \bibnamefont {Chang}}, \ and\ \bibinfo {author} {\bibfnamefont {F.~J.}\
  \bibnamefont {Garcia~de Abajo}},\ }\href {\doibase 10.1021/nl201771h}
  {\bibfield  {journal} {\bibinfo  {journal} {Nano Letters}\ }\textbf {\bibinfo
  {volume} {11}},\ \bibinfo {pages} {3370} (\bibinfo {year}
  {2011})}\BibitemShut {NoStop}%
\bibitem [{\citenamefont {Rodrigo}\ \emph {et~al.}(2015)\citenamefont
  {Rodrigo}, \citenamefont {Limaj}, \citenamefont {Janner}, \citenamefont
  {Etezadi}, \citenamefont {Garc{\'\i}a~de Abajo}, \citenamefont {Pruneri},\
  and\ \citenamefont {Altug}}]{Rodrigo15}%
  \BibitemOpen
  \bibfield  {author} {\bibinfo {author} {\bibfnamefont {D.}~\bibnamefont
  {Rodrigo}}, \bibinfo {author} {\bibfnamefont {O.}~\bibnamefont {Limaj}},
  \bibinfo {author} {\bibfnamefont {D.}~\bibnamefont {Janner}}, \bibinfo
  {author} {\bibfnamefont {D.}~\bibnamefont {Etezadi}}, \bibinfo {author}
  {\bibfnamefont {F.~J.}\ \bibnamefont {Garc{\'\i}a~de Abajo}}, \bibinfo
  {author} {\bibfnamefont {V.}~\bibnamefont {Pruneri}}, \ and\ \bibinfo
  {author} {\bibfnamefont {H.}~\bibnamefont {Altug}},\ }\href@noop {}
  {\bibfield  {journal} {\bibinfo  {journal} {Science}\ }\textbf {\bibinfo
  {volume} {349}},\ \bibinfo {pages} {165} (\bibinfo {year}
  {2015})}\BibitemShut {NoStop}%
\bibitem [{\citenamefont {Baqir}\ \emph {et~al.}(2019)\citenamefont {Baqir},
  \citenamefont {Choudhury}, \citenamefont {Farmani}, \citenamefont {Younas},
  \citenamefont {Arshad}, \citenamefont {Mir},\ and\ \citenamefont
  {Karimi}}]{Baqir19}%
  \BibitemOpen
  \bibfield  {author} {\bibinfo {author} {\bibfnamefont {M.~A.}\ \bibnamefont
  {Baqir}}, \bibinfo {author} {\bibfnamefont {P.~K.}\ \bibnamefont
  {Choudhury}}, \bibinfo {author} {\bibfnamefont {A.}~\bibnamefont {Farmani}},
  \bibinfo {author} {\bibfnamefont {T.}~\bibnamefont {Younas}}, \bibinfo
  {author} {\bibfnamefont {J.}~\bibnamefont {Arshad}}, \bibinfo {author}
  {\bibfnamefont {A.}~\bibnamefont {Mir}}, \ and\ \bibinfo {author}
  {\bibfnamefont {S.}~\bibnamefont {Karimi}},\ }\href {\doibase
  10.1109/JPHOT.2019.2931586} {\bibfield  {journal} {\bibinfo  {journal} {IEEE
  Photonics Journal}\ }\textbf {\bibinfo {volume} {11}},\ \bibinfo {pages} {1}
  (\bibinfo {year} {2019})}\BibitemShut {NoStop}%
\bibitem [{\citenamefont {Stauber}\ \emph {et~al.}(2020)\citenamefont
  {Stauber}, \citenamefont {Low},\ and\ \citenamefont
  {G{\'o}mez-Santos}}]{Stauber20}%
  \BibitemOpen
  \bibfield  {author} {\bibinfo {author} {\bibfnamefont {T.}~\bibnamefont
  {Stauber}}, \bibinfo {author} {\bibfnamefont {T.}~\bibnamefont {Low}}, \ and\
  \bibinfo {author} {\bibfnamefont {G.}~\bibnamefont {G{\'o}mez-Santos}},\
  }\href@noop {} {\bibfield  {journal} {\bibinfo  {journal} {Nano Letters}\
  }\textbf {\bibinfo {volume} {20}},\ \bibinfo {pages} {8711} (\bibinfo {year}
  {2020})}\BibitemShut {NoStop}%
\bibitem [{\citenamefont {Scholz}\ \emph {et~al.}(2013)\citenamefont {Scholz},
  \citenamefont {Stauber},\ and\ \citenamefont {Schliemann}}]{Scholz13}%
  \BibitemOpen
  \bibfield  {author} {\bibinfo {author} {\bibfnamefont {A.}~\bibnamefont
  {Scholz}}, \bibinfo {author} {\bibfnamefont {T.}~\bibnamefont {Stauber}}, \
  and\ \bibinfo {author} {\bibfnamefont {J.}~\bibnamefont {Schliemann}},\
  }\href {\doibase 10.1103/PhysRevB.88.035135} {\bibfield  {journal} {\bibinfo
  {journal} {Phys. Rev. B}\ }\textbf {\bibinfo {volume} {88}},\ \bibinfo
  {pages} {035135} (\bibinfo {year} {2013})}\BibitemShut {NoStop}%
\bibitem [{\citenamefont {Low}\ and\ \citenamefont {Avouris}(2014)}]{Low14}%
  \BibitemOpen
  \bibfield  {author} {\bibinfo {author} {\bibfnamefont {T.}~\bibnamefont
  {Low}}\ and\ \bibinfo {author} {\bibfnamefont {P.}~\bibnamefont {Avouris}},\
  }\href {\doibase 10.1021/nn406627u} {\bibfield  {journal} {\bibinfo
  {journal} {ACS Nano}\ }\textbf {\bibinfo {volume} {8}},\ \bibinfo {pages}
  {1086} (\bibinfo {year} {2014})}\BibitemShut {NoStop}%
\bibitem [{\citenamefont {Basov}\ \emph {et~al.}(2016)\citenamefont {Basov},
  \citenamefont {Fogler},\ and\ \citenamefont {Garc{\'\i}a~de
  Abajo}}]{Basov16}%
  \BibitemOpen
  \bibfield  {author} {\bibinfo {author} {\bibfnamefont {D.~N.}\ \bibnamefont
  {Basov}}, \bibinfo {author} {\bibfnamefont {M.~M.}\ \bibnamefont {Fogler}}, \
  and\ \bibinfo {author} {\bibfnamefont {F.~J.}\ \bibnamefont {Garc{\'\i}a~de
  Abajo}},\ }\href@noop {} {\bibfield  {journal} {\bibinfo  {journal}
  {Science}\ }\textbf {\bibinfo {volume} {354}} (\bibinfo {year}
  {2016})}\BibitemShut {NoStop}%
\bibitem [{\citenamefont {Mikhailov}\ and\ \citenamefont
  {Ziegler}(2007)}]{Mikhailov07}%
  \BibitemOpen
  \bibfield  {author} {\bibinfo {author} {\bibfnamefont {S.~A.}\ \bibnamefont
  {Mikhailov}}\ and\ \bibinfo {author} {\bibfnamefont {K.}~\bibnamefont
  {Ziegler}},\ }\href {\doibase 10.1103/PhysRevLett.99.016803} {\bibfield
  {journal} {\bibinfo  {journal} {Phys. Rev. Lett.}\ }\textbf {\bibinfo
  {volume} {99}},\ \bibinfo {pages} {016803} (\bibinfo {year}
  {2007})}\BibitemShut {NoStop}%
\bibitem [{\citenamefont {Ramos-Mendieta}\ \emph {et~al.}(2014)\citenamefont
  {Ramos-Mendieta}, \citenamefont {Hern\'andez-L\'opez},\ and\ \citenamefont
  {Palomino-Ovando}}]{Ramos-Mendieta14}%
  \BibitemOpen
  \bibfield  {author} {\bibinfo {author} {\bibfnamefont {F.}~\bibnamefont
  {Ramos-Mendieta}}, \bibinfo {author} {\bibfnamefont {J.~A.}\ \bibnamefont
  {Hern\'andez-L\'opez}}, \ and\ \bibinfo {author} {\bibfnamefont
  {M.}~\bibnamefont {Palomino-Ovando}},\ }\href@noop {} {\bibfield  {journal}
  {\bibinfo  {journal} {AIP Advances}\ }\textbf {\bibinfo {volume} {4}},\
  \bibinfo {pages} {067125} (\bibinfo {year} {2014})}\BibitemShut {NoStop}%
\bibitem [{\citenamefont {Menabde}\ \emph {et~al.}(2016)\citenamefont
  {Menabde}, \citenamefont {Mason}, \citenamefont {Kornev}, \citenamefont
  {Lee},\ and\ \citenamefont {Park}}]{Menabde16}%
  \BibitemOpen
  \bibfield  {author} {\bibinfo {author} {\bibfnamefont {S.~G.}\ \bibnamefont
  {Menabde}}, \bibinfo {author} {\bibfnamefont {D.~R.}\ \bibnamefont {Mason}},
  \bibinfo {author} {\bibfnamefont {E.~E.}\ \bibnamefont {Kornev}}, \bibinfo
  {author} {\bibfnamefont {C.}~\bibnamefont {Lee}}, \ and\ \bibinfo {author}
  {\bibfnamefont {N.}~\bibnamefont {Park}},\ }\href@noop {} {\bibfield
  {journal} {\bibinfo  {journal} {Scientific Reports}\ }\textbf {\bibinfo
  {volume} {6}},\ \bibinfo {pages} {21523} (\bibinfo {year}
  {2016})}\BibitemShut {NoStop}%
\bibitem [{\citenamefont {Zhang}\ \emph {et~al.}(2020)\citenamefont {Zhang},
  \citenamefont {Hu}, \citenamefont {Lin}, \citenamefont {Shen}, \citenamefont
  {Zhang},\ and\ \citenamefont {Chen}}]{Zhang20}%
  \BibitemOpen
  \bibfield  {author} {\bibinfo {author} {\bibfnamefont {X.}~\bibnamefont
  {Zhang}}, \bibinfo {author} {\bibfnamefont {H.}~\bibnamefont {Hu}}, \bibinfo
  {author} {\bibfnamefont {X.}~\bibnamefont {Lin}}, \bibinfo {author}
  {\bibfnamefont {L.}~\bibnamefont {Shen}}, \bibinfo {author} {\bibfnamefont
  {B.}~\bibnamefont {Zhang}}, \ and\ \bibinfo {author} {\bibfnamefont
  {H.}~\bibnamefont {Chen}},\ }\href {\doibase 10.1038/s41699-020-00159-z}
  {\bibfield  {journal} {\bibinfo  {journal} {npj 2D Materials and
  Applications}\ }\textbf {\bibinfo {volume} {4}},\ \bibinfo {pages} {25}
  (\bibinfo {year} {2020})}\BibitemShut {NoStop}%
\bibitem [{\citenamefont {Reserbat-Plantey}\ \emph {et~al.}(2021)\citenamefont
  {Reserbat-Plantey}, \citenamefont {Epstein}, \citenamefont {Torre},
  \citenamefont {Costa}, \citenamefont {Gon{\c c}alves}, \citenamefont
  {Mortensen}, \citenamefont {Polini}, \citenamefont {Song}, \citenamefont
  {Peres},\ and\ \citenamefont {Koppens}}]{Reserbat-Plantey21}%
  \BibitemOpen
  \bibfield  {author} {\bibinfo {author} {\bibfnamefont {A.}~\bibnamefont
  {Reserbat-Plantey}}, \bibinfo {author} {\bibfnamefont {I.}~\bibnamefont
  {Epstein}}, \bibinfo {author} {\bibfnamefont {I.}~\bibnamefont {Torre}},
  \bibinfo {author} {\bibfnamefont {A.~T.}\ \bibnamefont {Costa}}, \bibinfo
  {author} {\bibfnamefont {P.~A.~D.}\ \bibnamefont {Gon{\c c}alves}}, \bibinfo
  {author} {\bibfnamefont {N.~A.}\ \bibnamefont {Mortensen}}, \bibinfo {author}
  {\bibfnamefont {M.}~\bibnamefont {Polini}}, \bibinfo {author} {\bibfnamefont
  {J.~C.~W.}\ \bibnamefont {Song}}, \bibinfo {author} {\bibfnamefont
  {N.~M.~R.}\ \bibnamefont {Peres}}, \ and\ \bibinfo {author} {\bibfnamefont
  {F.~H.~L.}\ \bibnamefont {Koppens}},\ }\href@noop {} {\bibfield  {journal}
  {\bibinfo  {journal} {ACS Photonics}\ }\textbf {\bibinfo {volume} {8}},\
  \bibinfo {pages} {85} (\bibinfo {year} {2021})}\BibitemShut {NoStop}%
\bibitem [{\citenamefont {Chamanara}\ and\ \citenamefont
  {Caloz}(2016)}]{Chamanara16}%
  \BibitemOpen
  \bibfield  {author} {\bibinfo {author} {\bibfnamefont {N.}~\bibnamefont
  {Chamanara}}\ and\ \bibinfo {author} {\bibfnamefont {C.}~\bibnamefont
  {Caloz}},\ }\href {\doibase 10.1103/PhysRevB.94.075413} {\bibfield  {journal}
  {\bibinfo  {journal} {Phys. Rev. B}\ }\textbf {\bibinfo {volume} {94}},\
  \bibinfo {pages} {075413} (\bibinfo {year} {2016})}\BibitemShut {NoStop}%
\bibitem [{\citenamefont {Jablan}\ \emph {et~al.}(2011)\citenamefont {Jablan},
  \citenamefont {Buljan},\ and\ \citenamefont {Solja\v{c}i\'{c}}}]{Jablan11}%
  \BibitemOpen
  \bibfield  {author} {\bibinfo {author} {\bibfnamefont {M.}~\bibnamefont
  {Jablan}}, \bibinfo {author} {\bibfnamefont {H.}~\bibnamefont {Buljan}}, \
  and\ \bibinfo {author} {\bibfnamefont {M.}~\bibnamefont {Solja\v{c}i\'{c}}},\
  }\href {\doibase 10.1364/OE.19.011236} {\bibfield  {journal} {\bibinfo
  {journal} {Opt. Express}\ }\textbf {\bibinfo {volume} {19}},\ \bibinfo
  {pages} {11236} (\bibinfo {year} {2011})}\BibitemShut {NoStop}%
\bibitem [{\citenamefont {Andreeva}\ \emph {et~al.}(2018)\citenamefont
  {Andreeva}, \citenamefont {Luskin},\ and\ \citenamefont
  {Margetis}}]{Andreeva18}%
  \BibitemOpen
  \bibfield  {author} {\bibinfo {author} {\bibfnamefont {V.}~\bibnamefont
  {Andreeva}}, \bibinfo {author} {\bibfnamefont {M.}~\bibnamefont {Luskin}}, \
  and\ \bibinfo {author} {\bibfnamefont {D.}~\bibnamefont {Margetis}},\ }\href
  {\doibase 10.1103/PhysRevB.98.195407} {\bibfield  {journal} {\bibinfo
  {journal} {Phys. Rev. B}\ }\textbf {\bibinfo {volume} {98}},\ \bibinfo
  {pages} {195407} (\bibinfo {year} {2018})}\BibitemShut {NoStop}%
\bibitem [{\citenamefont {Principi}\ \emph {et~al.}(2009)\citenamefont
  {Principi}, \citenamefont {Polini},\ and\ \citenamefont
  {Vignale}}]{Principi09}%
  \BibitemOpen
  \bibfield  {author} {\bibinfo {author} {\bibfnamefont {A.}~\bibnamefont
  {Principi}}, \bibinfo {author} {\bibfnamefont {M.}~\bibnamefont {Polini}}, \
  and\ \bibinfo {author} {\bibfnamefont {G.}~\bibnamefont {Vignale}},\ }\href
  {\doibase 10.1103/PhysRevB.80.075418} {\bibfield  {journal} {\bibinfo
  {journal} {Phys. Rev. B}\ }\textbf {\bibinfo {volume} {80}},\ \bibinfo
  {pages} {075418} (\bibinfo {year} {2009})}\BibitemShut {NoStop}%
\bibitem [{\citenamefont {Stauber}\ and\ \citenamefont
  {G\'omez-Santos}(2010)}]{Stauber10}%
  \BibitemOpen
  \bibfield  {author} {\bibinfo {author} {\bibfnamefont {T.}~\bibnamefont
  {Stauber}}\ and\ \bibinfo {author} {\bibfnamefont {G.}~\bibnamefont
  {G\'omez-Santos}},\ }\href {\doibase 10.1103/PhysRevB.82.155412} {\bibfield
  {journal} {\bibinfo  {journal} {Phys. Rev. B}\ }\textbf {\bibinfo {volume}
  {82}},\ \bibinfo {pages} {155412} (\bibinfo {year} {2010})}\BibitemShut
  {NoStop}%
\bibitem [{\citenamefont {Guti\'errez-Rubio}\ \emph {et~al.}()\citenamefont
  {Guti\'errez-Rubio}, \citenamefont {Stauber},\ and\ \citenamefont
  {Guinea}}]{Angel13}%
  \BibitemOpen
  \bibfield  {author} {\bibinfo {author} {\bibfnamefont {A.}~\bibnamefont
  {Guti\'errez-Rubio}}, \bibinfo {author} {\bibfnamefont {T.}~\bibnamefont
  {Stauber}}, \ and\ \bibinfo {author} {\bibfnamefont {F.}~\bibnamefont
  {Guinea}},\ }\href@noop {} {\bibfield  {journal} {\bibinfo  {journal}
  {Journal of Optics}\ }\textbf {\bibinfo {volume} {15}},\ \bibinfo {pages}
  {114005}}\BibitemShut {NoStop}%
\bibitem [{\citenamefont {Basov}\ \emph {et~al.}(2021)\citenamefont {Basov},
  \citenamefont {Asenjo-Garcia}, \citenamefont {Schuck}, \citenamefont {Zhu},\
  and\ \citenamefont {Rubio}}]{Basov21}%
  \BibitemOpen
  \bibfield  {author} {\bibinfo {author} {\bibfnamefont {D.~N.}\ \bibnamefont
  {Basov}}, \bibinfo {author} {\bibfnamefont {A.}~\bibnamefont
  {Asenjo-Garcia}}, \bibinfo {author} {\bibfnamefont {P.~J.}\ \bibnamefont
  {Schuck}}, \bibinfo {author} {\bibfnamefont {X.}~\bibnamefont {Zhu}}, \ and\
  \bibinfo {author} {\bibfnamefont {A.}~\bibnamefont {Rubio}},\ }\href
  {\doibase doi:10.1515/nanoph-2020-0449} {\bibfield  {journal} {\bibinfo
  {journal} {Nanophotonics}\ }\textbf {\bibinfo {volume} {10}},\ \bibinfo
  {pages} {549} (\bibinfo {year} {2021})}\BibitemShut {NoStop}%
\bibitem [{\citenamefont {Stauber}\ and\ \citenamefont
  {G\'omez-Santos}(2012)}]{StauberPRB12}%
  \BibitemOpen
  \bibfield  {author} {\bibinfo {author} {\bibfnamefont {T.}~\bibnamefont
  {Stauber}}\ and\ \bibinfo {author} {\bibfnamefont {G.}~\bibnamefont
  {G\'omez-Santos}},\ }\href {\doibase 10.1103/PhysRevB.85.075410} {\bibfield
  {journal} {\bibinfo  {journal} {Phys. Rev. B}\ }\textbf {\bibinfo {volume}
  {85}},\ \bibinfo {pages} {075410} (\bibinfo {year} {2012})}\BibitemShut
  {NoStop}%
\bibitem [{\citenamefont {Stauber}\ \emph
  {et~al.}(2018{\natexlab{a}})\citenamefont {Stauber}, \citenamefont {Low},\
  and\ \citenamefont {G\'omez-Santos}}]{Stauber18}%
  \BibitemOpen
  \bibfield  {author} {\bibinfo {author} {\bibfnamefont {T.}~\bibnamefont
  {Stauber}}, \bibinfo {author} {\bibfnamefont {T.}~\bibnamefont {Low}}, \ and\
  \bibinfo {author} {\bibfnamefont {G.}~\bibnamefont {G\'omez-Santos}},\ }\href
  {\doibase 10.1103/PhysRevLett.120.046801} {\bibfield  {journal} {\bibinfo
  {journal} {Phys. Rev. Lett.}\ }\textbf {\bibinfo {volume} {120}},\ \bibinfo
  {pages} {046801} (\bibinfo {year} {2018}{\natexlab{a}})}\BibitemShut
  {NoStop}%
\bibitem [{\citenamefont {Stauber}\ and\ \citenamefont
  {G{\'{o}}mez-Santos}(2012)}]{Stauber12}%
  \BibitemOpen
  \bibfield  {author} {\bibinfo {author} {\bibfnamefont {T.}~\bibnamefont
  {Stauber}}\ and\ \bibinfo {author} {\bibfnamefont {G.}~\bibnamefont
  {G{\'{o}}mez-Santos}},\ }\href {\doibase 10.1088/1367-2630/14/10/105018}
  {\bibfield  {journal} {\bibinfo  {journal} {New Journal of Physics}\ }\textbf
  {\bibinfo {volume} {14}},\ \bibinfo {pages} {105018} (\bibinfo {year}
  {2012})}\BibitemShut {NoStop}%
\bibitem [{\citenamefont {Giuliani}\ and\ \citenamefont
  {Vignale}(2005)}]{giuliani05}%
  \BibitemOpen
  \bibfield  {author} {\bibinfo {author} {\bibfnamefont {G.}~\bibnamefont
  {Giuliani}}\ and\ \bibinfo {author} {\bibfnamefont {G.}~\bibnamefont
  {Vignale}},\ }\href {\doibase 10.1017/CBO9780511619915} {\emph {\bibinfo
  {title} {Quantum Theory of the Electron Liquid}}}\ (\bibinfo  {publisher}
  {Cambridge University Press},\ \bibinfo {year} {2005})\BibitemShut {NoStop}%
\bibitem [{\citenamefont {Kotov}\ \emph {et~al.}(2013)\citenamefont {Kotov},
  \citenamefont {Kol'chenko},\ and\ \citenamefont {Lozovik}}]{Kotov13}%
  \BibitemOpen
  \bibfield  {author} {\bibinfo {author} {\bibfnamefont {O.}~\bibnamefont
  {Kotov}}, \bibinfo {author} {\bibfnamefont {M.}~\bibnamefont {Kol'chenko}}, \
  and\ \bibinfo {author} {\bibfnamefont {Y.~E.}\ \bibnamefont {Lozovik}},\
  }\href {\doibase 10.1364/OE.21.013533} {\bibfield  {journal} {\bibinfo
  {journal} {Opt. Express}\ }\textbf {\bibinfo {volume} {21}},\ \bibinfo
  {pages} {13533} (\bibinfo {year} {2013})}\BibitemShut {NoStop}%
\bibitem [{\citenamefont {Castro~Neto}\ \emph {et~al.}(2009)\citenamefont
  {Castro~Neto}, \citenamefont {Guinea}, \citenamefont {Peres}, \citenamefont
  {Novoselov},\ and\ \citenamefont {Geim}}]{Castro09}%
  \BibitemOpen
  \bibfield  {author} {\bibinfo {author} {\bibfnamefont {A.~H.}\ \bibnamefont
  {Castro~Neto}}, \bibinfo {author} {\bibfnamefont {F.}~\bibnamefont {Guinea}},
  \bibinfo {author} {\bibfnamefont {N.~M.~R.}\ \bibnamefont {Peres}}, \bibinfo
  {author} {\bibfnamefont {K.~S.}\ \bibnamefont {Novoselov}}, \ and\ \bibinfo
  {author} {\bibfnamefont {A.~K.}\ \bibnamefont {Geim}},\ }\href {\doibase
  10.1103/RevModPhys.81.109} {\bibfield  {journal} {\bibinfo  {journal} {Rev.
  Mod. Phys.}\ }\textbf {\bibinfo {volume} {81}},\ \bibinfo {pages} {109}
  (\bibinfo {year} {2009})}\BibitemShut {NoStop}%
\bibitem [{\citenamefont {McCann}\ and\ \citenamefont
  {Koshino}(2013)}]{McCann13}%
  \BibitemOpen
  \bibfield  {author} {\bibinfo {author} {\bibfnamefont {E.}~\bibnamefont
  {McCann}}\ and\ \bibinfo {author} {\bibfnamefont {M.}~\bibnamefont
  {Koshino}},\ }\href {\doibase 10.1088/0034-4885/76/5/056503} {\bibfield
  {journal} {\bibinfo  {journal} {Reports on Progress in Physics}\ }\textbf
  {\bibinfo {volume} {76}},\ \bibinfo {pages} {056503} (\bibinfo {year}
  {2013})}\BibitemShut {NoStop}%
\bibitem [{\citenamefont {Rozhkov}\ \emph {et~al.}(2016)\citenamefont
  {Rozhkov}, \citenamefont {Sboychakov}, \citenamefont {Rakhmanov},\ and\
  \citenamefont {Nori}}]{Rozhkov16}%
  \BibitemOpen
  \bibfield  {author} {\bibinfo {author} {\bibfnamefont {A.}~\bibnamefont
  {Rozhkov}}, \bibinfo {author} {\bibfnamefont {A.}~\bibnamefont {Sboychakov}},
  \bibinfo {author} {\bibfnamefont {A.}~\bibnamefont {Rakhmanov}}, \ and\
  \bibinfo {author} {\bibfnamefont {F.}~\bibnamefont {Nori}},\ }\href {\doibase
  https://doi.org/10.1016/j.physrep.2016.07.003} {\bibfield  {journal}
  {\bibinfo  {journal} {Physics Reports}\ }\textbf {\bibinfo {volume} {648}},\
  \bibinfo {pages} {1} (\bibinfo {year} {2016})}\BibitemShut {NoStop}%
\bibitem [{\citenamefont {Tabert}\ and\ \citenamefont
  {Nicol}(2012)}]{Tabert12}%
  \BibitemOpen
  \bibfield  {author} {\bibinfo {author} {\bibfnamefont {C.~J.}\ \bibnamefont
  {Tabert}}\ and\ \bibinfo {author} {\bibfnamefont {E.~J.}\ \bibnamefont
  {Nicol}},\ }\href {\doibase 10.1103/PhysRevB.86.075439} {\bibfield  {journal}
  {\bibinfo  {journal} {Phys. Rev. B}\ }\textbf {\bibinfo {volume} {86}},\
  \bibinfo {pages} {075439} (\bibinfo {year} {2012})}\BibitemShut {NoStop}%
\bibitem [{\citenamefont {Rold\'an}\ and\ \citenamefont
  {Brey}(2013)}]{Roldan13}%
  \BibitemOpen
  \bibfield  {author} {\bibinfo {author} {\bibfnamefont {R.}~\bibnamefont
  {Rold\'an}}\ and\ \bibinfo {author} {\bibfnamefont {L.}~\bibnamefont
  {Brey}},\ }\href {\doibase 10.1103/PhysRevB.88.115420} {\bibfield  {journal}
  {\bibinfo  {journal} {Phys. Rev. B}\ }\textbf {\bibinfo {volume} {88}},\
  \bibinfo {pages} {115420} (\bibinfo {year} {2013})}\BibitemShut {NoStop}%
\bibitem [{\citenamefont {Abergel}\ and\ \citenamefont
  {Fal'ko}(2007)}]{Abergel07}%
  \BibitemOpen
  \bibfield  {author} {\bibinfo {author} {\bibfnamefont {D.~S.~L.}\
  \bibnamefont {Abergel}}\ and\ \bibinfo {author} {\bibfnamefont {V.~I.}\
  \bibnamefont {Fal'ko}},\ }\href {\doibase 10.1103/PhysRevB.75.155430}
  {\bibfield  {journal} {\bibinfo  {journal} {Phys. Rev. B}\ }\textbf {\bibinfo
  {volume} {75}},\ \bibinfo {pages} {155430} (\bibinfo {year}
  {2007})}\BibitemShut {NoStop}%
\bibitem [{\citenamefont {Low}\ \emph {et~al.}(2014)\citenamefont {Low},
  \citenamefont {Guinea}, \citenamefont {Yan}, \citenamefont {Xia},\ and\
  \citenamefont {Avouris}}]{Low14b}%
  \BibitemOpen
  \bibfield  {author} {\bibinfo {author} {\bibfnamefont {T.}~\bibnamefont
  {Low}}, \bibinfo {author} {\bibfnamefont {F.}~\bibnamefont {Guinea}},
  \bibinfo {author} {\bibfnamefont {H.}~\bibnamefont {Yan}}, \bibinfo {author}
  {\bibfnamefont {F.}~\bibnamefont {Xia}}, \ and\ \bibinfo {author}
  {\bibfnamefont {P.}~\bibnamefont {Avouris}},\ }\href {\doibase
  10.1103/PhysRevLett.112.116801} {\bibfield  {journal} {\bibinfo  {journal}
  {Phys. Rev. Lett.}\ }\textbf {\bibinfo {volume} {112}},\ \bibinfo {pages}
  {116801} (\bibinfo {year} {2014})}\BibitemShut {NoStop}%
\bibitem [{\citenamefont {Stauber}(2014{\natexlab{b}})}]{Stauber14b}%
  \BibitemOpen
  \bibfield  {author} {\bibinfo {author} {\bibfnamefont {T.}~\bibnamefont
  {Stauber}},\ }\href {\doibase 10.1088/0953-8984/26/12/123201} {\bibfield
  {journal} {\bibinfo  {journal} {Journal of Physics: Condensed Matter}\
  }\textbf {\bibinfo {volume} {26}},\ \bibinfo {pages} {123201} (\bibinfo
  {year} {2014}{\natexlab{b}})}\BibitemShut {NoStop}%
\bibitem [{\citenamefont {Nair}\ \emph {et~al.}(2008)\citenamefont {Nair},
  \citenamefont {Blake}, \citenamefont {Grigorenko}, \citenamefont {Novoselov},
  \citenamefont {Booth}, \citenamefont {Stauber}, \citenamefont {Peres},\ and\
  \citenamefont {Geim}}]{Nair08}%
  \BibitemOpen
  \bibfield  {author} {\bibinfo {author} {\bibfnamefont {R.~R.}\ \bibnamefont
  {Nair}}, \bibinfo {author} {\bibfnamefont {P.}~\bibnamefont {Blake}},
  \bibinfo {author} {\bibfnamefont {A.~N.}\ \bibnamefont {Grigorenko}},
  \bibinfo {author} {\bibfnamefont {K.~S.}\ \bibnamefont {Novoselov}}, \bibinfo
  {author} {\bibfnamefont {T.~J.}\ \bibnamefont {Booth}}, \bibinfo {author}
  {\bibfnamefont {T.}~\bibnamefont {Stauber}}, \bibinfo {author} {\bibfnamefont
  {N.~M.~R.}\ \bibnamefont {Peres}}, \ and\ \bibinfo {author} {\bibfnamefont
  {A.~K.}\ \bibnamefont {Geim}},\ }\href {\doibase 10.1126/science.1156965}
  {\bibfield  {journal} {\bibinfo  {journal} {Science}\ }\textbf {\bibinfo
  {volume} {320}},\ \bibinfo {pages} {1308} (\bibinfo {year}
  {2008})}\BibitemShut {NoStop}%
\bibitem [{\citenamefont {Nataf}\ \emph {et~al.}(2019)\citenamefont {Nataf},
  \citenamefont {Champel}, \citenamefont {Blatter},\ and\ \citenamefont
  {Basko}}]{Nataf19}%
  \BibitemOpen
  \bibfield  {author} {\bibinfo {author} {\bibfnamefont {P.}~\bibnamefont
  {Nataf}}, \bibinfo {author} {\bibfnamefont {T.}~\bibnamefont {Champel}},
  \bibinfo {author} {\bibfnamefont {G.}~\bibnamefont {Blatter}}, \ and\
  \bibinfo {author} {\bibfnamefont {D.~M.}\ \bibnamefont {Basko}},\ }\href
  {\doibase 10.1103/PhysRevLett.123.207402} {\bibfield  {journal} {\bibinfo
  {journal} {Phys. Rev. Lett.}\ }\textbf {\bibinfo {volume} {123}},\ \bibinfo
  {pages} {207402} (\bibinfo {year} {2019})}\BibitemShut {NoStop}%
\bibitem [{\citenamefont {Andolina}\ \emph {et~al.}(2020)\citenamefont
  {Andolina}, \citenamefont {Pellegrino}, \citenamefont {Giovannetti},
  \citenamefont {MacDonald},\ and\ \citenamefont {Polini}}]{Andolina20}%
  \BibitemOpen
  \bibfield  {author} {\bibinfo {author} {\bibfnamefont {G.~M.}\ \bibnamefont
  {Andolina}}, \bibinfo {author} {\bibfnamefont {F.~M.~D.}\ \bibnamefont
  {Pellegrino}}, \bibinfo {author} {\bibfnamefont {V.}~\bibnamefont
  {Giovannetti}}, \bibinfo {author} {\bibfnamefont {A.~H.}\ \bibnamefont
  {MacDonald}}, \ and\ \bibinfo {author} {\bibfnamefont {M.}~\bibnamefont
  {Polini}},\ }\href {\doibase 10.1103/PhysRevB.102.125137} {\bibfield
  {journal} {\bibinfo  {journal} {Phys. Rev. B}\ }\textbf {\bibinfo {volume}
  {102}},\ \bibinfo {pages} {125137} (\bibinfo {year} {2020})}\BibitemShut
  {NoStop}%
\bibitem [{\citenamefont {Stauber}\ \emph
  {et~al.}(2018{\natexlab{b}})\citenamefont {Stauber}, \citenamefont {Low},\
  and\ \citenamefont {G\'omez-Santos}}]{Stauber18b}%
  \BibitemOpen
  \bibfield  {author} {\bibinfo {author} {\bibfnamefont {T.}~\bibnamefont
  {Stauber}}, \bibinfo {author} {\bibfnamefont {T.}~\bibnamefont {Low}}, \ and\
  \bibinfo {author} {\bibfnamefont {G.}~\bibnamefont {G\'omez-Santos}},\ }\href
  {\doibase 10.1103/PhysRevB.98.195414} {\bibfield  {journal} {\bibinfo
  {journal} {Phys. Rev. B}\ }\textbf {\bibinfo {volume} {98}},\ \bibinfo
  {pages} {195414} (\bibinfo {year} {2018}{\natexlab{b}})}\BibitemShut
  {NoStop}%
\bibitem [{\citenamefont {Sharpe}\ \emph {et~al.}(2021)\citenamefont {Sharpe},
  \citenamefont {Fox}, \citenamefont {Barnard}, \citenamefont {Finney},
  \citenamefont {Watanabe}, \citenamefont {Taniguchi}, \citenamefont
  {Kastner},\ and\ \citenamefont {Goldhaber-Gordon}}]{Sharpe21}%
  \BibitemOpen
  \bibfield  {author} {\bibinfo {author} {\bibfnamefont {A.~L.}\ \bibnamefont
  {Sharpe}}, \bibinfo {author} {\bibfnamefont {E.~J.}\ \bibnamefont {Fox}},
  \bibinfo {author} {\bibfnamefont {A.~W.}\ \bibnamefont {Barnard}}, \bibinfo
  {author} {\bibfnamefont {J.}~\bibnamefont {Finney}}, \bibinfo {author}
  {\bibfnamefont {K.}~\bibnamefont {Watanabe}}, \bibinfo {author}
  {\bibfnamefont {T.}~\bibnamefont {Taniguchi}}, \bibinfo {author}
  {\bibfnamefont {M.~A.}\ \bibnamefont {Kastner}}, \ and\ \bibinfo {author}
  {\bibfnamefont {D.}~\bibnamefont {Goldhaber-Gordon}},\ }\href@noop {}
  {\bibfield  {journal} {\bibinfo  {journal} {Nano Letters}\ }\textbf {\bibinfo
  {volume} {21}},\ \bibinfo {pages} {4299} (\bibinfo {year}
  {2021})}\BibitemShut {NoStop}%
\bibitem [{\citenamefont {Guerci}\ \emph {et~al.}(2021)\citenamefont {Guerci},
  \citenamefont {Simon},\ and\ \citenamefont {Mora}}]{Guerci21}%
  \BibitemOpen
  \bibfield  {author} {\bibinfo {author} {\bibfnamefont {D.}~\bibnamefont
  {Guerci}}, \bibinfo {author} {\bibfnamefont {P.}~\bibnamefont {Simon}}, \
  and\ \bibinfo {author} {\bibfnamefont {C.}~\bibnamefont {Mora}},\ }\href
  {\doibase 10.1103/PhysRevB.103.224436} {\bibfield  {journal} {\bibinfo
  {journal} {Phys. Rev. B}\ }\textbf {\bibinfo {volume} {103}},\ \bibinfo
  {pages} {224436} (\bibinfo {year} {2021})}\BibitemShut {NoStop}%
\bibitem [{\citenamefont {Rakhmanov}\ \emph {et~al.}(2012)\citenamefont
  {Rakhmanov}, \citenamefont {Rozhkov}, \citenamefont {Sboychakov},\ and\
  \citenamefont {Nori}}]{Rakhmanov12}%
  \BibitemOpen
  \bibfield  {author} {\bibinfo {author} {\bibfnamefont {A.~L.}\ \bibnamefont
  {Rakhmanov}}, \bibinfo {author} {\bibfnamefont {A.~V.}\ \bibnamefont
  {Rozhkov}}, \bibinfo {author} {\bibfnamefont {A.~O.}\ \bibnamefont
  {Sboychakov}}, \ and\ \bibinfo {author} {\bibfnamefont {F.}~\bibnamefont
  {Nori}},\ }\href {\doibase 10.1103/PhysRevLett.109.206801} {\bibfield
  {journal} {\bibinfo  {journal} {Phys. Rev. Lett.}\ }\textbf {\bibinfo
  {volume} {109}},\ \bibinfo {pages} {206801} (\bibinfo {year}
  {2012})}\BibitemShut {NoStop}%
\bibitem [{\citenamefont {Brey}\ and\ \citenamefont {Fertig}(2013)}]{Brey13}%
  \BibitemOpen
  \bibfield  {author} {\bibinfo {author} {\bibfnamefont {L.}~\bibnamefont
  {Brey}}\ and\ \bibinfo {author} {\bibfnamefont {H.~A.}\ \bibnamefont
  {Fertig}},\ }\href {\doibase 10.1103/PhysRevB.87.115411} {\bibfield
  {journal} {\bibinfo  {journal} {Phys. Rev. B}\ }\textbf {\bibinfo {volume}
  {87}},\ \bibinfo {pages} {115411} (\bibinfo {year} {2013})}\BibitemShut
  {NoStop}%
\bibitem [{\citenamefont {de~la Peña}\ \emph {et~al.}(2014)\citenamefont
  {de~la Peña}, \citenamefont {Scherer},\ and\ \citenamefont
  {Honerkamp}}]{Pena14}%
  \BibitemOpen
  \bibfield  {author} {\bibinfo {author} {\bibfnamefont {D.~S.}\ \bibnamefont
  {de~la Peña}}, \bibinfo {author} {\bibfnamefont {M.~M.}\ \bibnamefont
  {Scherer}}, \ and\ \bibinfo {author} {\bibfnamefont {C.}~\bibnamefont
  {Honerkamp}},\ }\href@noop {} {\bibfield  {journal} {\bibinfo  {journal}
  {Annalen der Physik}\ }\textbf {\bibinfo {volume} {526}},\ \bibinfo {pages}
  {366} (\bibinfo {year} {2014})}\BibitemShut {NoStop}%
\end{thebibliography}
\end{document}